\documentclass[sigconf,nonacm]{acmart}
\usepackage[T1]{fontenc}
\usepackage{graphicx}
\usepackage{amsmath}
\usepackage[disable]{todonotes}
\usepackage{placeins}
\usepackage{tabularx}
\usepackage{booktabs}
\usepackage{adjustbox}
\usepackage{longtable}
\usepackage[normalem]{ulem}

\usepackage{xparse}
\usepackage{stfloats}
\usepackage{subcaption}
\usepackage{csquotes}
\usepackage{mathtools}
\hypersetup{hidelinks}
\usepackage[capitalize,noabbrev]{cleveref}  %

\usepackage{tikz}
\usepackage{pgfplots}
\pgfplotsset{compat=1.17}
\usetikzlibrary{arrows.meta,positioning,calc,fit,backgrounds,plotmarks}

\definecolor{inkPrimary}{HTML}{0B0B0B}
\definecolor{inkSecondary}{HTML}{52514E}
\definecolor{inkMuted}{HTML}{898781}
\definecolor{figGrid}{HTML}{E1E0D9}
\definecolor{figWash}{HTML}{F0EFEC}

\definecolor{huePQ}{HTML}{2A78D6}
\definecolor{hueKM}{HTML}{EB6834}
\definecolor{hueVQ}{HTML}{1BAF7A}
\definecolor{seqLight}{HTML}{CDE2FB}
\definecolor{seqMid}{HTML}{6DA7EC}
\definecolor{seqDark}{HTML}{1C5CAB}
\definecolor{bestMark}{HTML}{D1345B}  %

\pgfplotsset{
  docid axis/.style={
    scale only axis, width=0.73\linewidth, height=0.49\linewidth,
    axis x line*=bottom, axis y line*=left,
    axis line style={inkMuted, line width=0.4pt},
    tick style={inkMuted, line width=0.4pt}, major tick length=2pt, tick align=outside,
    tick label style={font=\scriptsize, text=inkSecondary, /pgf/number format/fixed},
    label style={font=\footnotesize, text=inkPrimary},
    xlabel style={yshift=1.5pt},
    ylabel style={at={(axis description cs:-0.18,0.5)}, anchor=south},
    ymajorgrids, grid style={figGrid, line width=0.4pt},
    legend style={font=\scriptsize, draw=none, fill=none, text=inkPrimary},
    every axis plot/.append style={line width=0.8pt},
    error bars/error bar style={line width=0.5pt, opacity=0.45},
    error bars/error mark=none,
  },
  famPQ/.style={color=huePQ, mark=*, mark size=1.7pt, mark options={solid, fill=huePQ, draw=white, line width=0.3pt}},
  famRK/.style={color=hueKM, mark=square*, mark size=1.65pt, mark options={solid, fill=hueKM, draw=white, line width=0.3pt}},
  famPRK/.style={color=hueKM, mark=diamond*, mark size=2.1pt, mark options={solid, fill=white, draw=hueKM, line width=0.7pt}},
  famRV/.style={color=hueVQ, mark=square*, mark size=1.65pt, mark options={solid, fill=hueVQ, draw=white, line width=0.3pt}},
  famPRV/.style={color=hueVQ, mark=diamond*, mark size=2.1pt, mark options={solid, fill=white, draw=hueVQ, line width=0.7pt}},
  dsMS/.style={solid},
  dsNQ/.style={dash pattern=on 2.4pt off 1.5pt},
  fit line/.style={inkMuted, solid, no markers, line width=0.6pt},
}

\DeclareMathOperator*{\argmin}{arg\,min}

\DeclarePairedDelimiter\abs{\lvert}{\rvert}      %
\DeclarePairedDelimiter\norm{\lVert}{\rVert}     %
\DeclarePairedDelimiter\set{\lbrace}{\rbrace}    %

\begin{document}

\title{A Systematic Study of Semantic ID Spaces for Generative Information Retrieval}
\author{Alexia Allal}
\email{alexia.allal@artefact.com}
\affiliation{\institution{Artefact Research Center}\city{Paris}\country{France}}
\author{Hicham Randrianarivo}
\email{hicham.randrianarivo@artefact.com}
\affiliation{\institution{Artefact Research Center}\city{Paris}\country{France}}
\author{Sylvain Lamprier}
\email{sylvain.lamprier@univ-angers.fr}
\affiliation{\institution{LERIA, Angers University}\city{Angers}\country{France}}
\renewcommand{\shortauthors}{Allal et al.}
\begin{abstract}
Generative Information Retrieval (GIR) has emerged as a transformative paradigm, shifting document retrieval from a traditional “retrieve-and-rank” workflow to sequence-to-sequence generation, where a model directly predicts document identifiers (DocIDs). While the semantic design of these DocIDs is known to be critical for performance, a fundamental question remains under-explored: what makes a good DocID? Current approaches rely heavily on computationally expensive downstream evaluations, hindering systematic analysis and rapid iteration. In this work, we address this challenge by presenting a comprehensive study on the properties, metrics, and trade-offs that define effective numerical DocIDs. Specifically, our contributions are threefold:
First, we propose a unified framework that unifies Product Quantization (PQ) and Residual Quantization (RQ), and their hybrid variants within a single design space. This enables us to systematically study key DocID properties, such as hierarchy versus parallelism, as well as the impact of hyperparameters like DocID length and codebook size.
Second, we define a suite of training-free, intrinsic metrics, to quantify DocID quality and evaluate structural fidelity without the overhead of full model training.
Through extensive experiments on MS MARCO 300K and NQ320K, we analyze how these structural properties influence retrieval effectiveness.
\end{abstract}
\maketitle

\section{Introduction}

Generative Information Retrieval (GIR)~\cite{tay2022transformermemorydifferentiablesearch} has emerged as a compelling paradigm, shifting document retrieval from a traditional "retrieve-and-rank" workflow to sequence-to-sequence generation, where a model directly predicts a discrete document identifier (DocID) given a query. 
The success of GIR hinges on DocID design: structuring documents under semantic codes induces a coarse-to-fine taxonomy, allowing the model to make high-level neighborhood decisions early in decoding~\cite{tay2022transformermemorydifferentiablesearch,wang2023neuralcorpusindexerdocument}. 

DocIDs can be either textual (e.g., titles, URLs, or n-grams) or numerical codes ranging from arbitrary identifiers (e.g., sequential counters or random integers) to structured semantic codes produced by vector quantization. Our work focuses on the latter, where the dominant paradigms are Product Quantization (PQ)~\cite{jegou2011productquantization}, which decomposes the embedding space into independent parallel subspaces quantized separately (PQ), and Residual Quantization (RQ)~\cite{Chen2010ApproximateNN}, which iteratively encodes a vector by quantizing the residual error of the preceding step. We study two RQ variants that differ in how the codebooks are trained: R-KMeans~\cite{deng2025onerecunifyingretrieverank, zeng2023scalableeffectivegenerativeinformation} fits each layer by $k$-means on the residuals of the previous one, and R-VQ~\cite{lee2022rqvae, rajput2023recommendersystemsgenerativeretrieval} learns all layers end-to-end on a reconstruction loss (\Cref{sec:background}).

Despite their prevalence, existing studies have three notable limitations. First, most papers fix a single quantization method and configuration of codebook vocabulary size $V$ and DocID length $M$~\cite{tay2022transformermemorydifferentiablesearch, wang2023neuralcorpusindexerdocument, Mekonnen_2025, zeng2023scalableeffectivegenerativeinformation}, leaving the design space unexplored. Second, parallel (PQ) and residual (RQ) schemes are treated as disjoint paradigms, overlooking hybrid configurations that combine parallel factorization with local residual depth. Finally, evaluating DocID quality relies almost exclusively on full end-to-end model training, treating identifier construction as an opaque preprocessing step without probing its underlying geometric and statistical properties.

We structure our investigation around three research questions. \textbf{RQ1:} How do the codebook size $V$, the DocID length $M$ and the codebook training affect downstream retrieval, and how do residual variants (R-KMeans, R-VQ) compare with PQ across the design space?
\textbf{RQ2:} Can a hybrid Product--Residual Quantization (PQ$\times$RQ) scheme bridge parallel factorization and residual hierarchy to improve retrieval trade-offs?
\textbf{RQ3:} Can training-free intrinsic metrics diagnose DocID structural viability prior to full model training?

Our contributions are fourfold. We propose a parameterized framework that unifies PQ, R-KMeans and R-VQ into a single family and exposes a regime not studied systematically for GIR, hybrid Product--Residual Quantization (PQ$\times$RQ), which lets us probe the trade-off between parallel factorization and hierarchical depth. Within this family, we train and evaluate generative retrieval models on $60$ MS MARCO 300K and $48$ Natural Questions (NQ) configurations, isolating the effect of the encoding method, codebook size, DocID length and codebook training. To diagnose DocID spaces before any training, we define a set of training-free intrinsic metrics: the Uniqueness Ratio $\mathcal{U}$, Normalized Entropy $\mathcal{H}$, Shared Semantic Similarity $\mathcal{S}$, and Contrastive Alignment Preservation $\mathcal{A}$.

\section{Background}
\label{sec:background}

\subsection{Generative Retrieval}
\label{sec:gr}
GIR~\cite{tay2022transformermemorydifferentiablesearch,zhuang2023bridginggapindexingretrieval} replaces traditional retrieve-and-rank pipelines by training a sequence-to-sequence model $P_\theta$ to directly generate the identifier (DocID) of a relevant document given a query.

Formally, let $\mathcal{X}=\{d^{(1)}, \dots, d^{(N)}\}$ be a document corpus and $\mathcal{Q}$ a query set, where each query $q \in \mathcal{Q}$ has a relevant subset $q(\mathcal{X}) \subseteq \mathcal{X}$. An encoding function $\phi : \mathcal{X} \rightarrow \Omega = \{1, \dots, V\}^M$ maps each document $d$ to a discrete DocID sequence $\phi(d) = \mathbf{y} = (y_1, \dots, y_M)$ of $M$ tokens, each from a codebook of size $V$. The autoregressive model minimizes token-level cross-entropy:

\begin{equation}
\label{eq:gr-loss}
    \mathcal{L} = -\sum_{q \in \mathcal{Q}} \sum_{d \in q(\mathcal{X})} \sum_{m=1}^M \log P_\theta(\phi(d)_m \mid \phi(d)_{<m}, q),
\end{equation}
where $\phi(d)_m$ denotes the $m$-th token of the code of document $d$, and $\phi(d)_{<m}$ denotes the preceding tokens. %

At inference, candidate DocIDs are generated via prefix tree-constrained beam search~\cite{decao2021autoregressiveentityretrieval} to guarantee valid sequences, which map deterministically back to documents via collection lookup.

\subsection{DocID Encoding}
Literature classifies DocIDs into content-derived strings, such as document titles, URLs~\cite{ren2023tome}, N-grams~\cite{zhang2024generative}, or pseudo-queries~\cite{tang2023semanticenhanceddifferentiablesearchindex, li2023multiviewidentifiersenhancedgenerative}, and computationally-generated numerical codes. While early numerical approaches used arbitrary integer strings~\cite{tay2022transformermemorydifferentiablesearch, pradeep2023doesgenerativeretrievalscale}, or hashing~\cite{salakhutdinov2009semantichashing}, modern methods apply vector quantization to preserve continuous semantic structure~\cite{Babenko2014AdditiveQF, Chen2010ApproximateNN, Ge2014OptimizedPQ, wang2014optimizedcartesiankmeans}.

In this setting, the encoder factorizes as $\phi=\phi^y \circ \phi^z$, where $\phi^z: \mathcal{X} \rightarrow \mathcal{Z} \equiv \mathbb{R}^h$ is a pre-trained embedder (e.g., DPR~\cite{karpukhin2020densepassageretrieval} or GTR~\cite{ni2021largedualencodersgeneralizable}) producing continuous representations $z^{(i)}$ of dimension $h$, and $\phi^y: \mathcal{Z} \rightarrow \Omega$ is a quantizer mapping $z^{(i)}$ to a discrete sequence $\mathbf{y}^{(i)}$.

\textbf{Product Quantization} (PQ)~\cite{jegou2011productquantization} splits $z$ into $C$ sub-vectors $z = z^1 \oplus \dots \oplus z^C$ of dimension $h/C$ and quantizes each one independently with its own codebook of $V$ codewords $\mathcal{C}^c = (v^c_1, \dots, v^c_V) \in \mathbb{R}^{V \times h/C}$:
\begin{equation}\label{eq:pq}
y^{c} = \argmin_{u} \norm{z^{c} - v^{c}_{u}}, \qquad c = 1, \dots, C,
\end{equation}
giving the DocID $\mathbf{y} = (y^1, \dots, y^C)$~\cite{zhou2022ultronultimateretrievercorpus}. Training each subspace codebook via offline $k$-means or a reconstruction loss is mathematically equivalent; we thus denote this single-layer baseline simply as PQ.

\textbf{Residual Quantization} (RQ)~\cite{Chen2010ApproximateNN, rajput2023recommendersystemsgenerativeretrieval} instead builds a coarse-to-fine DocID $\mathbf{y} = (y^1, \dots, y^L)$ of $L$ layers, as tree-based clustering did in early work~\cite{tay2022transformermemorydifferentiablesearch, wang2023neuralcorpusindexerdocument, pradeep2023doesgenerativeretrievalscale, zhuang2023bridginggapindexingretrieval}: starting from $z^{0} = z$, each layer $l$ quantizes the residual of the previous one with a codebook $\mathcal{C}^l = (v^l_1, \dots, v^l_V) \in \mathbb{R}^{V \times h}$:
\begin{equation}\label{eq:rq}
y^{l} = \argmin_{u} \norm{z^{l-1} - v^{l}_{u}}, \qquad z^{l} = z^{l-1} - v^{l}_{y^{l}}, \qquad l = 1, \dots, L.
\end{equation}
Residual codebooks are trained either offline, by sequential $k$-means on the residuals of each layer (R-KMeans)~\cite{damico2026deployingsemanticidbasedgenerative, deng2025onerecunifyingretrieverank, zeng2023scalableeffectivegenerativeinformation}, or end-to-end, by gradient descent on the reconstruction loss $\mathcal{L}_{\mathrm{RQ}}(z) = \sum_{l=1}^L \norm{z^{l-1}-v^l_{y^l}}^2$ (R-VQ)~\cite{lee2022rqvae, esser2021tamingtransformershighresolutionimage, ju2025generativerecommendationsemanticids}.

\section{Semantic ID Spaces Study through a Unified Framework}

\subsection{Unifying Parallelism and Hierarchy: The Product--Residual Continuum}
\label{sec:unified-framework}

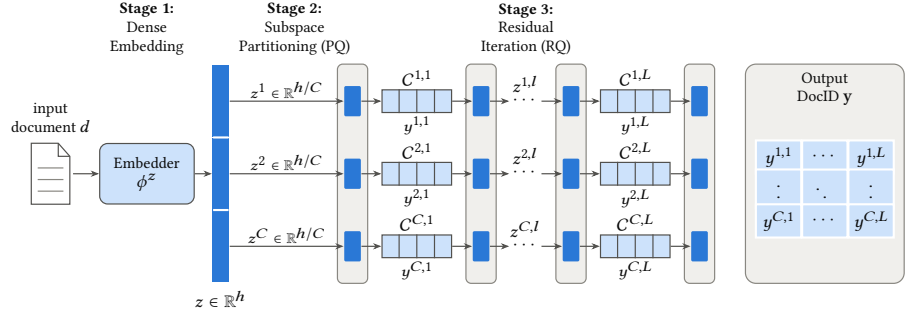
\begin{figure*}[tp]
    \centering
    \begin{tikzpicture}[x=0.85cm, y=0.96cm, font=\scriptsize, text=inkPrimary, >={Stealth[length=3.2pt]},
  every path/.style={draw=inkSecondary, line width=0.45pt},
  code/.style={draw=none, fill=huePQ, rounded corners=0.6pt, minimum width=0.22cm, minimum height=0.40cm, inner sep=0pt},
  column/.style={draw=inkMuted, fill=figWash, rounded corners=2pt, minimum width=0.40cm, minimum height=2.9cm, inner sep=0pt},
  stage/.style={font=\scriptsize, align=center, anchor=south, inner sep=1pt}]
\newcommand{\codebook}[3]{%
  \node[draw=inkSecondary, fill=seqLight, minimum width=0.92cm, minimum height=0.30cm, inner sep=0pt] (#1) at #2 {};
  \foreach \k in {1,2,3} \draw[draw=inkSecondary] ($(#1.south west)!\k/4!(#1.south east)$) -- ($(#1.north west)!\k/4!(#1.north east)$);
  \node[anchor=south, inner sep=1pt] at (#1.north) {#3};}
\draw (0,-0.42) -- (0,0.42) -- (0.42,0.42) -- (0.6,0.24) -- (0.6,-0.42) -- cycle; \draw (0.42,0.42) -- (0.42,0.24) -- (0.6,0.24);
\foreach \yy in {0.08,-0.08,-0.24} \draw[draw=inkMuted] (0.1,\yy) -- (0.5,\yy);
\node[anchor=south, align=center] at (0.3,0.46) {input\\document $d$};
\node[draw=inkSecondary, fill=seqLight, rounded corners=3pt, minimum width=1.25cm, minimum height=0.8cm, align=center] (emb) at (1.8,0) {Embedder\\$\phi^z$};
\draw[->] (0.65,0) -- (emb.west);
\foreach \yy/\c in {1.0/1, 0/2, -1.0/C} \node[code, minimum height=0.94cm, rounded corners=0pt] (z\c) at (2.95,\yy) {};
\foreach \yy in {0.5,-0.5} \draw[draw=white, line width=0.8pt] (2.83,\yy) -- (3.07,\yy);
\draw[->] (emb.east) -- (z2.west);
\node[anchor=north] at (2.95,-1.5) {$z \in \mathbb{R}^{h}$};
\foreach \x/\n in {5.0/a, 7.0/b, 8.4/c, 10.4/e} \node[column] (col\n) at (\x,0) {};
\foreach \yy/\c in {1.0/1, 0/2, -1.0/C} {
  \foreach \x/\n in {5.0/a, 7.0/b, 8.4/c, 10.4/e} \node[code] (\n\c) at (\x,\yy) {};
  \draw[->] (z\c.east |- 0,\yy) -- node[above, inner sep=1pt, font=\tiny, pos=0.5] {$z^{\c} \in \mathbb{R}^{h/C}$} (a\c.west);
  \codebook{b1\c}{(6.0,\yy)}{$\mathcal{C}^{\c,1}$}
  \codebook{bL\c}{(9.4,\yy)}{$\mathcal{C}^{\c,L}$}
  \draw[->] (a\c.east) -- (b1\c.west); \draw[->] (b1\c.east) -- (b\c.west);
  \node[below, inner sep=1pt, font=\tiny] at (b1\c.south) {$y^{\c,1}$};
  \draw[->] (b\c.east) -- ++(0.3,0) node[right, inner sep=1pt] (dots\c) {$\cdots$};
  \node[anchor=south, inner sep=1pt] at (dots\c.north) {$z^{\c,l}$};
  \draw[->] (dots\c.east) -- (c\c.west);
  \draw[->] (c\c.east) -- (bL\c.west); \draw[->] (bL\c.east) -- (e\c.west);
  \node[below, inner sep=1pt, font=\tiny] at (bL\c.south) {$y^{\c,L}$};
}
\node[draw=inkMuted, fill=figWash, rounded corners=4pt, minimum width=2.1cm, minimum height=2.9cm] (out) at (12.35,0) {};
\node[anchor=north, align=center] at (out.north) {Output\\DocID $\mathbf{y}$};
\begin{scope}[shift={(11.63,-0.65)}]
  \foreach \i/\j/\txt in {0/2/{y^{1,1}}, 1/2/{\cdots}, 2/2/{y^{1,L}}, 0/1/{\vdots}, 1/1/{\ddots}, 2/1/{\vdots}, 0/0/{y^{C,1}}, 1/0/{\cdots}, 2/0/{y^{C,L}}}
    \node[draw=white, line width=0.8pt, fill=seqLight, minimum width=0.6cm, minimum height=0.44cm, inner sep=0pt] at (\i*0.72,\j*0.44) {$\txt$};
\end{scope}
\node[stage] at (1.8,1.6) {\textbf{Stage 1:}\\Dense\\Embedding};
\node[stage] at (4.1,1.6) {\textbf{Stage 2:}\\Subspace\\Partitioning (PQ)};
\node[stage] at (7.7,1.6) {\textbf{Stage 3:}\\Residual\\Iteration (RQ)};
\end{tikzpicture}
    \caption{Unified framework for semantic DocID encoding. Stage 1: the embedder $\phi^z$ maps a document $d$ to a dense vector $z$. Stage 2 splits $z$ into $C$ sub-vectors $z^c$ (PQ, $C>1$). Stage 3 quantizes each sub-vector with $L$ residual codebooks $\mathcal{C}^{c,l}$ of $V$ codewords (RQ, $L>1$), each step emitting one token $y^{c,l}$ of the DocID $\mathbf{y}$ ($M = C \times L$ tokens). $(C, L)$ and the codebook training ($k$-means or learned) recover PQ ($L=1$), R-KMeans and R-VQ ($C=1$), and the hybrids PQ$\times$R-KMeans and PQ$\times$R-VQ ($C, L > 1$).}
    \label{fig:unified-framework}
\end{figure*}

We formalize semantic DocID construction by modeling horizontal decomposition (PQ-style) and vertical refinement (RQ-style) as limiting cases of a single discrete encoding operator $\phi^y_{\lambda}$ ($\lambda = (C, L, V)$). Depicted in \Cref{fig:unified-framework}, an input embedding $z \in \mathbb{R}^h$ is first split into $C$ parallel sub-vectors $z^c \in \mathbb{R}^{h/C}$. Each sub-vector $z^c$ then undergoes an $L$-step Residual Quantization (RQ) process (\Cref{sec:background}), with residuals $z^{c,l}$ and codebooks $\mathcal{C}^{c,l} = (v^{c,l}_1, \dots, v^{c,l}_V) \in \mathbb{R}^{V \times h/C}$ ($c \le C$, $l \le L$). Codebooks are trained either via sequential $k$-means on residual vectors (PQ$\times$R-KMeans) or end-to-end gradient minimization (PQ$\times$R-VQ).

With $z^{c,0} = z^c$, each sub-vector is quantized by the residual recursion
\begin{equation}\label{eq:prq-matrix}
y^{c,l} = \argmin_{u} \norm{z^{c,l-1} - v^{c,l}_u}, \qquad z^{c,l} = z^{c,l-1} - v^{c,l}_{y^{c,l}},
\end{equation}
giving the DocID $\mathbf{y} = (y^{c,l})_{c \le C,\, l \le L}$ of $M = C \times L$ tokens (\Cref{fig:unified-framework}); $C=1$ gives RQ~\eqref{eq:rq} and $L=1$ gives PQ~\eqref{eq:pq} with $y^c = y^{c,1}$.
The autoregressive decoder consumes $\mathbf{y}$ flattened column-by-column into $(y_1, \dots, y_M)$ to preserve hierarchy. The interior ($C, L > 1$) is the hybrid PQ$\times$RQ regime. Unlike OneSearch~\cite{chen2026onesearch}, which applies PQ only to the final residual of a global RQ, and PRQ-KMeans~\cite{luo2026prqkmeans}, a purely residual tokenizer despite its name, PQ$\times$RQ splits $z$ upfront into $C$ sub-spaces with an $L$-step hierarchy within each, so that the model navigates parallel hierarchies simultaneously.

\subsection{Intrinsic Structural Properties}
\label{sec:intrinsic-metrics}
Evaluating DocIDs solely through downstream generative model training is computationally expensive and leaves a blind spot regarding their structural integrity.

To circumvent this, intrinsic evaluation of discrete codebooks has recently emerged as a crucial diagnostic paradigm in generative recommendation~\cite{chen2026makesgoodsemanticid}. Extending this perspective to GIR, we assess the structural viability of candidate DocID spaces prior to sequence-to-sequence training with the metrics below.

\subsubsection{Distribution and Balance Metrics}
\label{sec:distribution-metrics}
To prevent training instability caused by highly skewed document assignments (where many documents collapse into few clusters), we evaluate the usage of the identifier space.

\paragraph{Uniqueness Ratio $\mathcal{U}$}
In GIR, when distinct documents are assigned the exact same DocID, an \textit{identifier collision} occurs, mechanically capping maximum achievable recall. We define the Uniqueness Ratio $\mathcal{U} \in (0, 1]$ simply as the proportion of unique DocID sequences generated across the corpus of $N$ documents. While $\mathcal{U}=1$ is the intuitive goal, documents with identical texts always share a DocID, so $\mathcal{U}$ cannot exceed the fraction of distinct texts, $\mathcal{U}_{max}$.

\paragraph{Normalized Entropy $\mathcal{H}$} 
To measure the uniformity of token assignments, we compute the Shannon entropy across individual token positions. For position $m \in \{1, \dots, M\}$, let $\mathcal{V}_m$ be the set of codewords $u$ used at position $m$, and $P_m(u)$ the fraction of the $N$ documents with $y_m = u$. Normalized Entropy $\mathcal{H}$ is the mean normalized entropy across all $M$ positions:

\begin{equation}
\label{eq:entropy}
\mathcal{H} = \frac{1}{M} \sum_{m=1}^{M} \mathcal{H}_m, \quad \text{where } \mathcal{H}_m = \frac{-\sum_{u \in \mathcal{V}_m} P_m(u) \log P_m(u)}{\log \abs{\mathcal{V}_m}}.
\end{equation}

A value of $\mathcal{H} = 1.0$ indicates that the tokens used at a position are used equally often; unused codebook entries do not lower $\mathcal{H}$.

\subsubsection{Shared Semantic Similarity $\mathcal{S}$}
\label{subsec:shared-similarity}
To evaluate local semantic coherence, Shared Semantic Similarity ($\mathcal{S}$) measures the average continuous embedding similarity between documents sharing identical discrete identifiers under a budget of $k \in \{1, \dots, M\}$ unmasked tokens. While any DocID representation $\mathbf{y} \in \mathbb{R}^{C \times L}$ ($M = C \times L$) can be evaluated under any paradigm, the budget masking function $f_k(\mathbf{y})$ adapts to three structural assumptions:
\begin{itemize}
    \item \textbf{Parallel (suited for PQ):} $f_k$ retains a random subset of $k$ distinct positions, ignoring positional hierarchy.
    \item \textbf{Hierarchical (suited for RQ):} Assuming a single sequence ($C=1, L=M$), $f_k$ retains the strictly contiguous prefix of length $k$.
    \item \textbf{Adapted (generalized hybrid):} Distributes $k$ tokens across $C$ parallel subspaces ($\sum k_c = k, k_c \le L$), retaining the hierarchical prefix of length $k_c$ in each. Note that \textit{Adapted} generalizes both paradigms, reducing to \textit{Parallel} when $L=1$ (PQ) and to \textit{Hierarchical} when $C=1$ (RQ).
\end{itemize}

Given a masking function $f_k$, $\mathcal{S}(k)$ is the conditional expectation of cosine similarity $S_c$ between documents sharing identical masked representations:
\begin{equation}
\label{eq:shared-sim}
    \mathcal{S}(k) = \mathbb{E}\Big[ S_c(z^{(i)}, z^{(j)}) \;\Big|\; f_k(\mathbf{y}^{(i)}) = f_k(\mathbf{y}^{(j)}) \Big].
\end{equation}
We report the global mean $\mathcal{S}$ averaged over all budget levels $k$.

\subsubsection{Contrastive Alignment Preservation $\mathcal{A}$}
Contrastive Alignment Preservation ($\mathcal{A}$) measures how faithfully discrete DocID distances preserve relative continuous embedding proximities. We first define a parameterized discrete similarity operator $S_d^{(\hat{C}, \hat{L})}$ between two identifier matrices $\mathbf{y}^{(i)}, \mathbf{y}^{(j)}$ over evaluated subspace dimensions $\hat{C}$ and depth levels $\hat{L}$:

\begin{equation}
\label{eq:sd-param}
    S_d^{(\hat{C}, \hat{L})}(\mathbf{y}^{(i)}, \mathbf{y}^{(j)}) = \frac{1}{M} \sum_{c=1}^{\hat{C}} \sum_{l=1}^{\hat{L}} \mathbf{1}\set{\mathbf{y}^{(i)}_{c, 1:l} = \mathbf{y}^{(j)}_{c, 1:l}},
\end{equation}
where $\mathbf{y}_{c, 1:l}$ denotes the prefix up to level $l$ within subspace $c$. Setting $(\hat{C}, \hat{L})$ recovers the metric variants corresponding to our structural paradigms:

\begin{itemize}
    \item \textbf{Hierarchical ($S_d^{\text{hier}}$):} Setting $\hat{C}=1, \hat{L}=M$ evaluates the normalized longest continuous common prefix, making it strictly adapted to purely hierarchical DocIDs (RQ).
    \item \textbf{Parallel ($S_d^{\text{parallel}}$):} Setting $\hat{C}=M, \hat{L}=1$ measures the unranked proportion of matching tokens across independent subspaces, tailored to flat parallel DocIDs (PQ).
    \item \textbf{Adapted ($S_d^{\text{adapted}}$):} Setting $\hat{C}=C, \hat{L}=L$ computes the average hierarchical prefix length across all $C$ parallel subspaces, yielding a hybrid metric natively suited for Product-Residual spaces (PQxRQ).
\end{itemize}

We sample document triplets $(d^{(i)}, d^{(j)}, d^{(k)})$ from the corpus, discarding tied states ($S_d(\mathbf{y}^{(i)}, \mathbf{y}^{(j)}) = S_d(\mathbf{y}^{(i)}, \mathbf{y}^{(k)})$) to build an evaluation set of $T=1{,}000$ valid triplets. The alignment rate $\mathcal{A}$ measures the probability that relative continuous embedding similarity $S_c$ aligns with discrete DocID ordering $S_d$:

\begin{equation}
\label{eq:alignment}
\mathcal{A} = P\big(S_c(z^{(i)}, z^{(j)}) > S_c(z^{(i)}, z^{(k)}) \,\big|\, S_d(\mathbf{y}^{(i)}, \mathbf{y}^{(j)}) > S_d(\mathbf{y}^{(i)}, \mathbf{y}^{(k)})\big).
\end{equation}
A higher $\mathcal{A}$ confirms that discrete DocID paths faithfully preserve local relative ordering.

\section{Results}

\subsection{Experimental Setup}
\label{sec:experimental-setup}

\paragraph{Training pipeline}
We follow a decoupled two-stage setup based on the DDRO codebase~\cite{Mekonnen_2025}. First, static document embeddings are generated using a frozen \texttt{gtr-t5-base}~\cite{ni2021largedualencodersgeneralizable} ($h=768$) and quantized into DocIDs. Second, a sequence-to-sequence model is trained by supervised fine-tuning only, in DDRO's three supervised steps (document, pseudo-query, then query to DocID), without its pairwise relevance optimization~\cite{Mekonnen_2025}.

\paragraph{Datasets}
We evaluate on MS MARCO 300K~\cite{nguyen2016msmarco}, processed as in DDRO and NCI~\cite{Mekonnen_2025,wang2023neuralcorpusindexerdocument} ($319{,}927$ documents with a positive query) and Natural Questions (NQ320K, hereafter NQ)~\cite{kwiatkowski-etal-2019-natural,wang2023neuralcorpusindexerdocument} ($109{,}739$ documents after deduplication). Following the strict DDRO protocol~\cite{Mekonnen_2025}, NQ is deduplicated by title with a strict one-to-one query-document mapping and disjoint train/eval document splits.

\paragraph{Configurations}

We evaluate codebook sizes $V \in \{128, 256, 512\}$ and structural dimensions $C, L \in \{1, 2, 4, 6, 8, 12, 16, 24\}$, testing hybrid PQ$\times$RQ configurations at $V=256$ across sequence budgets $M = C \times L \in [4, 48]$. R-VQ codebooks are initialized via $k$-means and optimized for $100$ epochs using AdamW to minimize the residual reconstruction loss (\Cref{eq:rq}). For generative retrieval, we fine-tune a \texttt{t5-small}~\cite{raffel2020exploringlimits} model (batch size 128, AdamW, learning rate $10^{-3}$ with linear decay) and report Recall@1, Recall@10 and MRR@10 on the MS MARCO dev queries and the $1{,}713$ NQ evaluation queries. Decoding uses prefix-constrained beam search ($10$ beams for validation during training, $100$ for all reported results); models are trained for $10$, $20$ and $10$ epochs over the three stages. 
We release the code, DocIDs and datasets.\footnote{Released upon acceptance.}

\subsection{Downstream Performance along the Parallelism vs. Hierarchy Continuum}

\subsubsection{Main Method Comparison}

\begin{figure*}[tp]
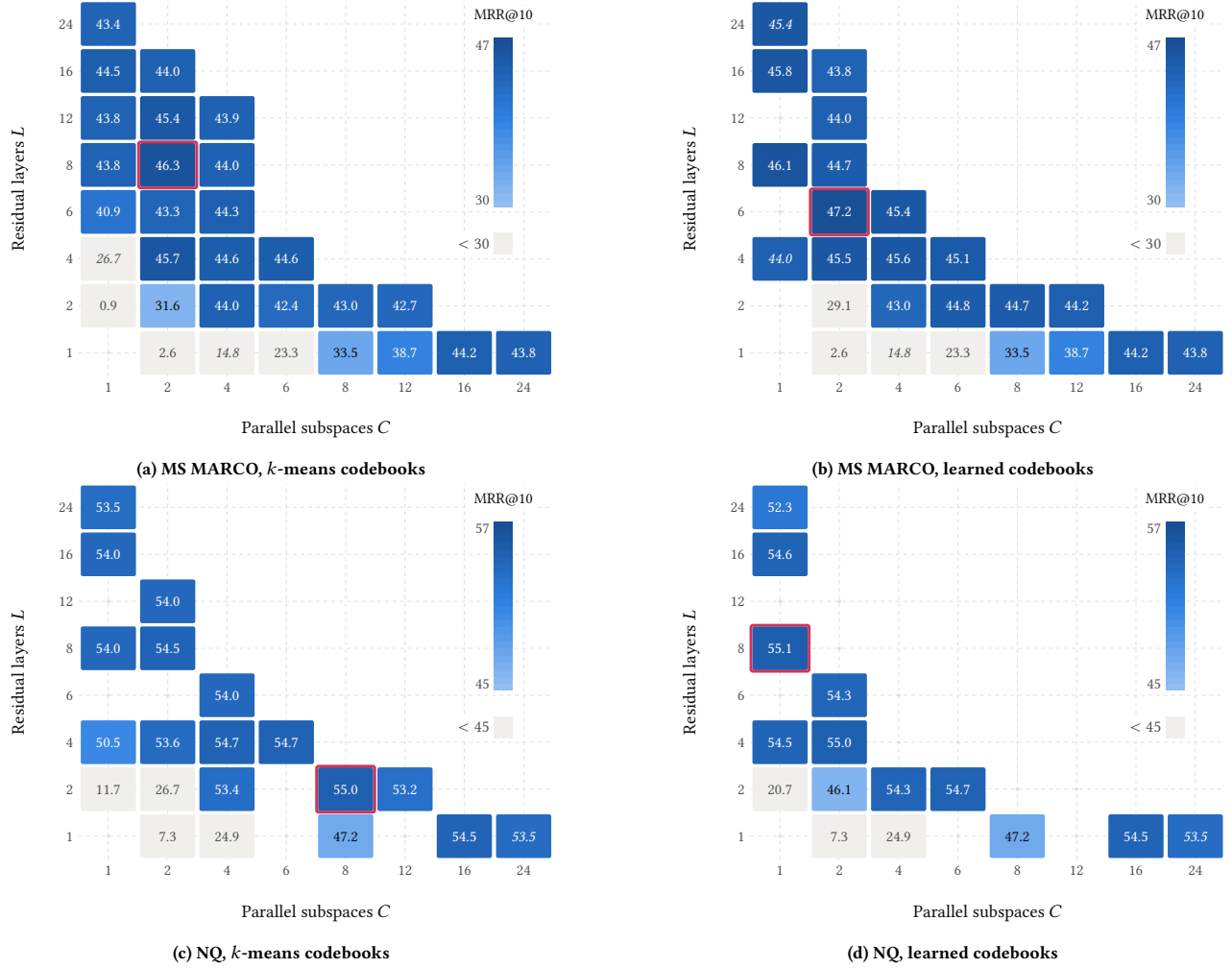

  \centering
  \begin{subfigure}[t]{0.48\linewidth}
    \centering
    \input{figures/tikz/v3_cl_grid_msmarco_kmeans}
    \caption{MS MARCO, $k$-means codebooks}
    \label{fig:cl-sub1}
  \end{subfigure}
  \hfill
  \begin{subfigure}[t]{0.48\linewidth}
    \centering
    \input{figures/tikz/v3_cl_grid_msmarco_vq}
    \caption{MS MARCO, learned codebooks}
    \label{fig:cl-sub2}
  \end{subfigure}

  \begin{subfigure}[t]{0.48\linewidth}
    \centering
    \begin{tikzpicture}[x=\linewidth/10.45, y=\linewidth/13.2, font=\scriptsize, text=inkSecondary]
\draw[figGrid, line width=0.3pt, dash pattern=on 1.5pt off 1.5pt] (0.500,0) -- (0.500,8.000); \draw[figGrid, line width=0.3pt, dash pattern=on 1.5pt off 1.5pt] (0,0.500) -- (8.000,0.500);
\draw[figGrid, line width=0.3pt, dash pattern=on 1.5pt off 1.5pt] (1.500,0) -- (1.500,8.000); \draw[figGrid, line width=0.3pt, dash pattern=on 1.5pt off 1.5pt] (0,1.500) -- (8.000,1.500);
\draw[figGrid, line width=0.3pt, dash pattern=on 1.5pt off 1.5pt] (2.500,0) -- (2.500,8.000); \draw[figGrid, line width=0.3pt, dash pattern=on 1.5pt off 1.5pt] (0,2.500) -- (8.000,2.500);
\draw[figGrid, line width=0.3pt, dash pattern=on 1.5pt off 1.5pt] (3.500,0) -- (3.500,8.000); \draw[figGrid, line width=0.3pt, dash pattern=on 1.5pt off 1.5pt] (0,3.500) -- (8.000,3.500);
\draw[figGrid, line width=0.3pt, dash pattern=on 1.5pt off 1.5pt] (4.500,0) -- (4.500,8.000); \draw[figGrid, line width=0.3pt, dash pattern=on 1.5pt off 1.5pt] (0,4.500) -- (8.000,4.500);
\draw[figGrid, line width=0.3pt, dash pattern=on 1.5pt off 1.5pt] (5.500,0) -- (5.500,8.000); \draw[figGrid, line width=0.3pt, dash pattern=on 1.5pt off 1.5pt] (0,5.500) -- (8.000,5.500);
\draw[figGrid, line width=0.3pt, dash pattern=on 1.5pt off 1.5pt] (6.500,0) -- (6.500,8.000); \draw[figGrid, line width=0.3pt, dash pattern=on 1.5pt off 1.5pt] (0,6.500) -- (8.000,6.500);
\draw[figGrid, line width=0.3pt, dash pattern=on 1.5pt off 1.5pt] (7.500,0) -- (7.500,8.000); \draw[figGrid, line width=0.3pt, dash pattern=on 1.5pt off 1.5pt] (0,7.500) -- (8.000,7.500);
\fill[figGrid] (0.500,0.500) circle[radius=0.7pt];
\fill[rounded corners=0.8pt, fill=figWash] (0.035,1.035) rectangle (0.965,1.965) node[midway, text=inkSecondary, font=\fontsize{6}{6}\selectfont, inner sep=0pt] {11.7};
\fill[rounded corners=0.8pt, fill={rgb,255:red,58;green,136;blue,229}] (0.035,2.035) rectangle (0.965,2.965) node[midway, text=white, font=\fontsize{6}{6}\selectfont, inner sep=0pt] {50.5};
\fill[figGrid] (0.500,3.500) circle[radius=0.7pt];
\fill[rounded corners=0.8pt, fill={rgb,255:red,36;green,104;blue,188}] (0.035,4.035) rectangle (0.965,4.965) node[midway, text=white, font=\fontsize{6}{6}\selectfont, inner sep=0pt] {54.0};
\fill[figGrid] (0.500,5.500) circle[radius=0.7pt];
\fill[rounded corners=0.8pt, fill={rgb,255:red,36;green,104;blue,188}] (0.035,6.035) rectangle (0.965,6.965) node[midway, text=white, font=\fontsize{6}{6}\selectfont, inner sep=0pt] {54.0};
\fill[rounded corners=0.8pt, fill={rgb,255:red,38;green,109;blue,196}] (0.035,7.035) rectangle (0.965,7.965) node[midway, text=white, font=\fontsize{6}{6}\selectfont, inner sep=0pt] {53.5};
\fill[rounded corners=0.8pt, fill=figWash] (1.035,0.035) rectangle (1.965,0.965) node[midway, text=inkSecondary, font=\fontsize{6}{6}\selectfont, inner sep=0pt] {7.3};
\fill[rounded corners=0.8pt, fill=figWash] (1.035,1.035) rectangle (1.965,1.965) node[midway, text=inkSecondary, font=\fontsize{6}{6}\selectfont, inner sep=0pt] {26.7};
\fill[rounded corners=0.8pt, fill={rgb,255:red,37;green,107;blue,193}] (1.035,2.035) rectangle (1.965,2.965) node[midway, text=white, font=\fontsize{6}{6}\selectfont, inner sep=0pt] {53.6};
\fill[figGrid] (1.500,3.500) circle[radius=0.7pt];
\fill[rounded corners=0.8pt, fill={rgb,255:red,33;green,100;blue,183}] (1.035,4.035) rectangle (1.965,4.965) node[midway, text=white, font=\fontsize{6}{6}\selectfont, inner sep=0pt] {54.5};
\fill[rounded corners=0.8pt, fill={rgb,255:red,36;green,104;blue,189}] (1.035,5.035) rectangle (1.965,5.965) node[midway, text=white, font=\fontsize{6}{6}\selectfont, inner sep=0pt] {54.0};
\fill[rounded corners=0.8pt, fill=figWash] (2.035,0.035) rectangle (2.965,0.965) node[midway, text=inkSecondary, font=\fontsize{6}{6}\selectfont, inner sep=0pt] {24.9};
\fill[rounded corners=0.8pt, fill={rgb,255:red,38;green,109;blue,197}] (2.035,1.035) rectangle (2.965,1.965) node[midway, text=white, font=\fontsize{6}{6}\selectfont, inner sep=0pt] {53.4};
\fill[rounded corners=0.8pt, fill={rgb,255:red,32;green,99;blue,180}] (2.035,2.035) rectangle (2.965,2.965) node[midway, text=white, font=\fontsize{6}{6}\selectfont, inner sep=0pt] {54.7};
\fill[rounded corners=0.8pt, fill={rgb,255:red,36;green,104;blue,189}] (2.035,3.035) rectangle (2.965,3.965) node[midway, text=white, font=\fontsize{6}{6}\selectfont, inner sep=0pt] {54.0};
\fill[figGrid] (3.500,0.500) circle[radius=0.7pt];
\fill[figGrid] (3.500,1.500) circle[radius=0.7pt];
\fill[rounded corners=0.8pt, fill={rgb,255:red,32;green,98;blue,180}] (3.035,2.035) rectangle (3.965,2.965) node[midway, text=white, font=\fontsize{6}{6}\selectfont, inner sep=0pt] {54.7};
\fill[rounded corners=0.8pt, fill={rgb,255:red,112;green,169;blue,236}] (4.035,0.035) rectangle (4.965,0.965) node[midway, text=inkPrimary, font=\fontsize{6}{6}\selectfont, inner sep=0pt] {47.2};
\fill[rounded corners=0.8pt, fill={rgb,255:red,30;green,95;blue,176}] (4.035,1.035) rectangle (4.965,1.965) node[midway, text=white, font=\fontsize{6}{6}\selectfont, inner sep=0pt] {55.0};
\draw[bestMark, line width=1.1pt, rounded corners=0.8pt] (4.010,1.010) rectangle (4.990,1.990);
\fill[figGrid] (5.500,0.500) circle[radius=0.7pt];
\fill[rounded corners=0.8pt, fill={rgb,255:red,39;green,111;blue,200}] (5.035,1.035) rectangle (5.965,1.965) node[midway, text=white, font=\fontsize{6}{6}\selectfont, inner sep=0pt] {53.2};
\fill[rounded corners=0.8pt, fill={rgb,255:red,33;green,100;blue,182}] (6.035,0.035) rectangle (6.965,0.965) node[midway, text=white, font=\fontsize{6}{6}\selectfont, inner sep=0pt] {54.5};
\fill[rounded corners=0.8pt, fill={rgb,255:red,38;green,108;blue,195}] (7.035,0.035) rectangle (7.965,0.965) node[midway, text=white, font=\fontsize{6}{6}\selectfont, inner sep=0pt] {\textit{53.5}};
\node[anchor=north, inner sep=2pt] at (0.500,0) {1};
\node[anchor=east, inner sep=2pt] at (0,0.500) {1};
\node[anchor=north, inner sep=2pt] at (1.500,0) {2};
\node[anchor=east, inner sep=2pt] at (0,1.500) {2};
\node[anchor=north, inner sep=2pt] at (2.500,0) {4};
\node[anchor=east, inner sep=2pt] at (0,2.500) {4};
\node[anchor=north, inner sep=2pt] at (3.500,0) {6};
\node[anchor=east, inner sep=2pt] at (0,3.500) {6};
\node[anchor=north, inner sep=2pt] at (4.500,0) {8};
\node[anchor=east, inner sep=2pt] at (0,4.500) {8};
\node[anchor=north, inner sep=2pt] at (5.500,0) {12};
\node[anchor=east, inner sep=2pt] at (0,5.500) {12};
\node[anchor=north, inner sep=2pt] at (6.500,0) {16};
\node[anchor=east, inner sep=2pt] at (0,6.500) {16};
\node[anchor=north, inner sep=2pt] at (7.500,0) {24};
\node[anchor=east, inner sep=2pt] at (0,7.500) {24};
\node[anchor=north, font=\footnotesize, text=inkPrimary] at (4.000,-0.78) {Parallel subspaces $C$};
\node[rotate=90, anchor=south, font=\footnotesize, text=inkPrimary] at (-0.72,4.000) {Residual layers $L$};
\fill[fill={rgb,255:red,146;green,190;blue,242}] (7.000,3.600) rectangle (7.320,3.755);
\fill[fill={rgb,255:red,138;green,185;blue,240}] (7.000,3.750) rectangle (7.320,3.905);
\fill[fill={rgb,255:red,130;green,180;blue,239}] (7.000,3.900) rectangle (7.320,4.055);
\fill[fill={rgb,255:red,122;green,175;blue,238}] (7.000,4.050) rectangle (7.320,4.205);
\fill[fill={rgb,255:red,114;green,170;blue,237}] (7.000,4.200) rectangle (7.320,4.355);
\fill[fill={rgb,255:red,106;green,165;blue,235}] (7.000,4.350) rectangle (7.320,4.505);
\fill[fill={rgb,255:red,98;green,160;blue,234}] (7.000,4.500) rectangle (7.320,4.655);
\fill[fill={rgb,255:red,90;green,155;blue,232}] (7.000,4.650) rectangle (7.320,4.805);
\fill[fill={rgb,255:red,82;green,150;blue,231}] (7.000,4.800) rectangle (7.320,4.955);
\fill[fill={rgb,255:red,73;green,145;blue,230}] (7.000,4.950) rectangle (7.320,5.105);
\fill[fill={rgb,255:red,64;green,139;blue,229}] (7.000,5.100) rectangle (7.320,5.255);
\fill[fill={rgb,255:red,56;green,134;blue,228}] (7.000,5.250) rectangle (7.320,5.405);
\fill[fill={rgb,255:red,51;green,129;blue,223}] (7.000,5.400) rectangle (7.320,5.555);
\fill[fill={rgb,255:red,46;green,124;blue,218}] (7.000,5.550) rectangle (7.320,5.705);
\fill[fill={rgb,255:red,42;green,119;blue,212}] (7.000,5.700) rectangle (7.320,5.855);
\fill[fill={rgb,255:red,40;green,115;blue,205}] (7.000,5.850) rectangle (7.320,6.005);
\fill[fill={rgb,255:red,38;green,110;blue,198}] (7.000,6.000) rectangle (7.320,6.155);
\fill[fill={rgb,255:red,37;green,105;blue,190}] (7.000,6.150) rectangle (7.320,6.305);
\fill[fill={rgb,255:red,34;green,101;blue,184}] (7.000,6.300) rectangle (7.320,6.455);
\fill[fill={rgb,255:red,31;green,96;blue,177}] (7.000,6.450) rectangle (7.320,6.605);
\fill[fill={rgb,255:red,28;green,92;blue,171}] (7.000,6.600) rectangle (7.320,6.755);
\fill[fill={rgb,255:red,27;green,87;blue,163}] (7.000,6.750) rectangle (7.320,6.905);
\fill[fill={rgb,255:red,25;green,83;blue,156}] (7.000,6.900) rectangle (7.320,7.055);
\fill[fill={rgb,255:red,24;green,79;blue,149}] (7.000,7.050) rectangle (7.320,7.205);
\fill[figWash] (7.000,2.600) rectangle (7.320,3.050);
\node[anchor=east, font=\fontsize{6}{6}\selectfont, inner sep=1.5pt, fill=white] at (7.000,7.050) {57};
\node[anchor=east, font=\fontsize{6}{6}\selectfont, inner sep=1.5pt, fill=white] at (7.000,3.750) {45};
\node[anchor=east, font=\fontsize{6}{6}\selectfont, inner sep=1.5pt, fill=white] at (7.000,2.825) {$<45$};
\node[anchor=south, font=\fontsize{6}{6}\selectfont, text=inkPrimary, inner sep=1.5pt, fill=white] at (7.160,7.450) {MRR@10};
\end{tikzpicture}
    \caption{NQ, $k$-means codebooks}
    \label{fig:cl-sub3}
    \vspace{-5pt}
  \end{subfigure}
  \hfill
  \begin{subfigure}[t]{0.48\linewidth}
    \centering
    \begin{tikzpicture}[x=\linewidth/10.45, y=\linewidth/13.2, font=\scriptsize, text=inkSecondary]
\draw[figGrid, line width=0.3pt, dash pattern=on 1.5pt off 1.5pt] (0.500,0) -- (0.500,8.000); \draw[figGrid, line width=0.3pt, dash pattern=on 1.5pt off 1.5pt] (0,0.500) -- (8.000,0.500);
\draw[figGrid, line width=0.3pt, dash pattern=on 1.5pt off 1.5pt] (1.500,0) -- (1.500,8.000); \draw[figGrid, line width=0.3pt, dash pattern=on 1.5pt off 1.5pt] (0,1.500) -- (8.000,1.500);
\draw[figGrid, line width=0.3pt, dash pattern=on 1.5pt off 1.5pt] (2.500,0) -- (2.500,8.000); \draw[figGrid, line width=0.3pt, dash pattern=on 1.5pt off 1.5pt] (0,2.500) -- (8.000,2.500);
\draw[figGrid, line width=0.3pt, dash pattern=on 1.5pt off 1.5pt] (3.500,0) -- (3.500,8.000); \draw[figGrid, line width=0.3pt, dash pattern=on 1.5pt off 1.5pt] (0,3.500) -- (8.000,3.500);
\draw[figGrid, line width=0.3pt, dash pattern=on 1.5pt off 1.5pt] (4.500,0) -- (4.500,8.000); \draw[figGrid, line width=0.3pt, dash pattern=on 1.5pt off 1.5pt] (0,4.500) -- (8.000,4.500);
\draw[figGrid, line width=0.3pt, dash pattern=on 1.5pt off 1.5pt] (5.500,0) -- (5.500,8.000); \draw[figGrid, line width=0.3pt, dash pattern=on 1.5pt off 1.5pt] (0,5.500) -- (8.000,5.500);
\draw[figGrid, line width=0.3pt, dash pattern=on 1.5pt off 1.5pt] (6.500,0) -- (6.500,8.000); \draw[figGrid, line width=0.3pt, dash pattern=on 1.5pt off 1.5pt] (0,6.500) -- (8.000,6.500);
\draw[figGrid, line width=0.3pt, dash pattern=on 1.5pt off 1.5pt] (7.500,0) -- (7.500,8.000); \draw[figGrid, line width=0.3pt, dash pattern=on 1.5pt off 1.5pt] (0,7.500) -- (8.000,7.500);
\fill[figGrid] (0.500,0.500) circle[radius=0.7pt];
\fill[rounded corners=0.8pt, fill=figWash] (0.035,1.035) rectangle (0.965,1.965) node[midway, text=inkSecondary, font=\fontsize{6}{6}\selectfont, inner sep=0pt] {20.7};
\fill[rounded corners=0.8pt, fill={rgb,255:red,33;green,100;blue,182}] (0.035,2.035) rectangle (0.965,2.965) node[midway, text=white, font=\fontsize{6}{6}\selectfont, inner sep=0pt] {54.5};
\fill[figGrid] (0.500,3.500) circle[radius=0.7pt];
\fill[rounded corners=0.8pt, fill={rgb,255:red,29;green,94;blue,174}] (0.035,4.035) rectangle (0.965,4.965) node[midway, text=white, font=\fontsize{6}{6}\selectfont, inner sep=0pt] {55.1};
\draw[bestMark, line width=1.1pt, rounded corners=0.8pt] (0.010,4.010) rectangle (0.990,4.990);
\fill[figGrid] (0.500,5.500) circle[radius=0.7pt];
\fill[rounded corners=0.8pt, fill={rgb,255:red,33;green,99;blue,181}] (0.035,6.035) rectangle (0.965,6.965) node[midway, text=white, font=\fontsize{6}{6}\selectfont, inner sep=0pt] {54.6};
\fill[rounded corners=0.8pt, fill={rgb,255:red,42;green,119;blue,212}] (0.035,7.035) rectangle (0.965,7.965) node[midway, text=white, font=\fontsize{6}{6}\selectfont, inner sep=0pt] {52.3};
\fill[rounded corners=0.8pt, fill=figWash] (1.035,0.035) rectangle (1.965,0.965) node[midway, text=inkSecondary, font=\fontsize{6}{6}\selectfont, inner sep=0pt] {7.3};
\fill[rounded corners=0.8pt, fill={rgb,255:red,130;green,180;blue,239}] (1.035,1.035) rectangle (1.965,1.965) node[midway, text=inkPrimary, font=\fontsize{6}{6}\selectfont, inner sep=0pt] {46.1};
\fill[rounded corners=0.8pt, fill={rgb,255:red,30;green,96;blue,176}] (1.035,2.035) rectangle (1.965,2.965) node[midway, text=white, font=\fontsize{6}{6}\selectfont, inner sep=0pt] {55.0};
\fill[rounded corners=0.8pt, fill={rgb,255:red,34;green,102;blue,185}] (1.035,3.035) rectangle (1.965,3.965) node[midway, text=white, font=\fontsize{6}{6}\selectfont, inner sep=0pt] {54.3};
\fill[figGrid] (1.500,4.500) circle[radius=0.7pt];
\fill[figGrid] (1.500,5.500) circle[radius=0.7pt];
\fill[rounded corners=0.8pt, fill=figWash] (2.035,0.035) rectangle (2.965,0.965) node[midway, text=inkSecondary, font=\fontsize{6}{6}\selectfont, inner sep=0pt] {24.9};
\fill[rounded corners=0.8pt, fill={rgb,255:red,34;green,101;blue,184}] (2.035,1.035) rectangle (2.965,1.965) node[midway, text=white, font=\fontsize{6}{6}\selectfont, inner sep=0pt] {54.3};
\fill[figGrid] (2.500,2.500) circle[radius=0.7pt];
\fill[figGrid] (2.500,3.500) circle[radius=0.7pt];
\fill[figGrid] (3.500,0.500) circle[radius=0.7pt];
\fill[rounded corners=0.8pt, fill={rgb,255:red,32;green,98;blue,180}] (3.035,1.035) rectangle (3.965,1.965) node[midway, text=white, font=\fontsize{6}{6}\selectfont, inner sep=0pt] {54.7};
\fill[figGrid] (3.500,2.500) circle[radius=0.7pt];
\fill[rounded corners=0.8pt, fill={rgb,255:red,112;green,169;blue,236}] (4.035,0.035) rectangle (4.965,0.965) node[midway, text=inkPrimary, font=\fontsize{6}{6}\selectfont, inner sep=0pt] {47.2};
\fill[figGrid] (4.500,1.500) circle[radius=0.7pt];
\fill[figGrid] (5.500,0.500) circle[radius=0.7pt];
\fill[figGrid] (5.500,1.500) circle[radius=0.7pt];
\fill[rounded corners=0.8pt, fill={rgb,255:red,33;green,100;blue,182}] (6.035,0.035) rectangle (6.965,0.965) node[midway, text=white, font=\fontsize{6}{6}\selectfont, inner sep=0pt] {54.5};
\fill[rounded corners=0.8pt, fill={rgb,255:red,38;green,108;blue,195}] (7.035,0.035) rectangle (7.965,0.965) node[midway, text=white, font=\fontsize{6}{6}\selectfont, inner sep=0pt] {\textit{53.5}};
\node[anchor=north, inner sep=2pt] at (0.500,0) {1};
\node[anchor=east, inner sep=2pt] at (0,0.500) {1};
\node[anchor=north, inner sep=2pt] at (1.500,0) {2};
\node[anchor=east, inner sep=2pt] at (0,1.500) {2};
\node[anchor=north, inner sep=2pt] at (2.500,0) {4};
\node[anchor=east, inner sep=2pt] at (0,2.500) {4};
\node[anchor=north, inner sep=2pt] at (3.500,0) {6};
\node[anchor=east, inner sep=2pt] at (0,3.500) {6};
\node[anchor=north, inner sep=2pt] at (4.500,0) {8};
\node[anchor=east, inner sep=2pt] at (0,4.500) {8};
\node[anchor=north, inner sep=2pt] at (5.500,0) {12};
\node[anchor=east, inner sep=2pt] at (0,5.500) {12};
\node[anchor=north, inner sep=2pt] at (6.500,0) {16};
\node[anchor=east, inner sep=2pt] at (0,6.500) {16};
\node[anchor=north, inner sep=2pt] at (7.500,0) {24};
\node[anchor=east, inner sep=2pt] at (0,7.500) {24};
\node[anchor=north, font=\footnotesize, text=inkPrimary] at (4.000,-0.78) {Parallel subspaces $C$};
\node[rotate=90, anchor=south, font=\footnotesize, text=inkPrimary] at (-0.72,4.000) {Residual layers $L$};
\fill[fill={rgb,255:red,146;green,190;blue,242}] (7.000,3.600) rectangle (7.320,3.755);
\fill[fill={rgb,255:red,138;green,185;blue,240}] (7.000,3.750) rectangle (7.320,3.905);
\fill[fill={rgb,255:red,130;green,180;blue,239}] (7.000,3.900) rectangle (7.320,4.055);
\fill[fill={rgb,255:red,122;green,175;blue,238}] (7.000,4.050) rectangle (7.320,4.205);
\fill[fill={rgb,255:red,114;green,170;blue,237}] (7.000,4.200) rectangle (7.320,4.355);
\fill[fill={rgb,255:red,106;green,165;blue,235}] (7.000,4.350) rectangle (7.320,4.505);
\fill[fill={rgb,255:red,98;green,160;blue,234}] (7.000,4.500) rectangle (7.320,4.655);
\fill[fill={rgb,255:red,90;green,155;blue,232}] (7.000,4.650) rectangle (7.320,4.805);
\fill[fill={rgb,255:red,82;green,150;blue,231}] (7.000,4.800) rectangle (7.320,4.955);
\fill[fill={rgb,255:red,73;green,145;blue,230}] (7.000,4.950) rectangle (7.320,5.105);
\fill[fill={rgb,255:red,64;green,139;blue,229}] (7.000,5.100) rectangle (7.320,5.255);
\fill[fill={rgb,255:red,56;green,134;blue,228}] (7.000,5.250) rectangle (7.320,5.405);
\fill[fill={rgb,255:red,51;green,129;blue,223}] (7.000,5.400) rectangle (7.320,5.555);
\fill[fill={rgb,255:red,46;green,124;blue,218}] (7.000,5.550) rectangle (7.320,5.705);
\fill[fill={rgb,255:red,42;green,119;blue,212}] (7.000,5.700) rectangle (7.320,5.855);
\fill[fill={rgb,255:red,40;green,115;blue,205}] (7.000,5.850) rectangle (7.320,6.005);
\fill[fill={rgb,255:red,38;green,110;blue,198}] (7.000,6.000) rectangle (7.320,6.155);
\fill[fill={rgb,255:red,37;green,105;blue,190}] (7.000,6.150) rectangle (7.320,6.305);
\fill[fill={rgb,255:red,34;green,101;blue,184}] (7.000,6.300) rectangle (7.320,6.455);
\fill[fill={rgb,255:red,31;green,96;blue,177}] (7.000,6.450) rectangle (7.320,6.605);
\fill[fill={rgb,255:red,28;green,92;blue,171}] (7.000,6.600) rectangle (7.320,6.755);
\fill[fill={rgb,255:red,27;green,87;blue,163}] (7.000,6.750) rectangle (7.320,6.905);
\fill[fill={rgb,255:red,25;green,83;blue,156}] (7.000,6.900) rectangle (7.320,7.055);
\fill[fill={rgb,255:red,24;green,79;blue,149}] (7.000,7.050) rectangle (7.320,7.205);
\fill[figWash] (7.000,2.600) rectangle (7.320,3.050);
\node[anchor=east, font=\fontsize{6}{6}\selectfont, inner sep=1.5pt, fill=white] at (7.000,7.050) {57};
\node[anchor=east, font=\fontsize{6}{6}\selectfont, inner sep=1.5pt, fill=white] at (7.000,3.750) {45};
\node[anchor=east, font=\fontsize{6}{6}\selectfont, inner sep=1.5pt, fill=white] at (7.000,2.825) {$<45$};
\node[anchor=south, font=\fontsize{6}{6}\selectfont, text=inkPrimary, inner sep=1.5pt, fill=white] at (7.160,7.450) {MRR@10};
\end{tikzpicture}
    \caption{NQ, learned codebooks}
    \label{fig:cl-sub4}
    \vspace{-5pt}
  \end{subfigure}

  \caption{MRR@10 (\%) over the grid of $C$ parallel subspaces and $L$ residual layers, with $V=256$ codewords per codebook. PQ has only one training regime: its row ($L=1$) is the same in every panel. Red outline: best cell of the panel; pale cells lie below the color range; italics: no run at $V=256$, mean over $V \in \{128, 512\}$.}
  \label{fig:cl-design-space}
  \vspace{-10pt}
\end{figure*}

The best configurations of the five families lie within $2.2$ MRR@10 points on both datasets (\Cref{fig:cl-design-space}; RQ1). While PQ favors large parallel factorization ($C=16$), pure hierarchical residual schemes (R-KMeans, R-VQ) score above it in Recall@10 and MRR@10 on both MS MARCO and NQ, and R-VQ reaches the best NQ result ($56.64\%$). With $k$-means codebooks, the hybrid PQ$\times$R-KMeans reaches $46.30\%$ MRR@10 on MS MARCO, the best result of the study, and $55.04\%$ on NQ by combining coarse parallel subspaces with local residual depth, above pure R-KMeans ($45.47\%$ and $54.70\%$).

\subsubsection{Codebook Size and Document Encoder}

Downstream retrieval is stable across codebook sizes: at fixed family and $M$, varying $V$ between $128$ and $512$ moves MRR@10 by $1.06$ (MS MARCO) and $1.53$ (NQ) points. Retrieval is equally stable across document encoders: with R-KMeans ($1{\times}16{\times}512$) on MS MARCO, five encoders (\texttt{gtr-t5-base}, \texttt{-large}, \texttt{-xl}, \texttt{bge-base-en-v1.5}, \texttt{ModernBERT-base}) keep MRR@10 between $42.31\%$ and $45.39\%$, with the standard \texttt{gtr-t5-base} highest.

\subsubsection{Exploring the Design Space: Sequence Length $M$ and Hybrid Configurations}

To analyze how structural choices shape downstream effectiveness, \Cref{fig:cl-design-space} maps the empirical performance surface across the entire $(C, L)$ design grid, for $V=256$. 
The total sequence length $M = C \times L$ acts as the primary driver of performance variance, but resilience to sequence compression varies significantly across quantization paradigms. 
PQ exhibits steep performance degradation once $C$ falls below $16$ ($38.70\%$ MRR@10 at $C=12$ on MS MARCO, $47.23\%$ at $C=8$ on NQ), as factorizing vectors into uncoordinated parallel chunks forces sub-tokens to compress too much information. 

Conversely, hierarchical residual schemes demonstrate substantially greater structural robustness: offline R-KMeans holds $43.80\%$ at $L=8$ but drops to $40.90\%$ at $L=6$ on MS MARCO, while end-to-end trained R-VQ remains remarkably resilient even under extreme compression ($L=4$). Across all schemes, NQ maintains acceptable retrieval performance at shorter sequence budgets than MS MARCO, consistent with its smaller corpus.

The interior of the grid ($C>1, L>1$) holds the hybrid spaces and answers RQ2. At their best, hybrids reach the top of the design space: PQ$\times$R-KMeans peaks at $46.3\%$ MRR@10 ($C=2, L=8$), the best MS MARCO result (\Cref{fig:cl-design-space}). Under tight budgets, residual depth is what preserves effectiveness: at $M=4$ on MS MARCO, R-VQ ($L=4$, $V=512$) reaches $44.0\%$, against $31.6\%$ and $26.2\%$ for PQ$\times$R-KMeans and PQ$\times$R-VQ ($C=L=2$), in line with its higher uniqueness ($\mathcal{U}=0.95$, against $0.66$ for PQ$\times$R-VQ).

\subsection{Intrinsic DocID Metric Analysis}
\label{sec:docid-analysis}

\begin{figure*}[!ht]
  \centering
  \begin{tikzpicture}
\begin{axis}[hide axis, scale only axis, width=1pt, height=1pt, xmin=0, xmax=1, ymin=0, ymax=1,
  legend columns=-1, legend style={draw=none, fill=none, font=\scriptsize, text=inkPrimary, at={(0.5,0.5)}, anchor=center,
  /tikz/every even column/.append style={column sep=6pt}}]
\addlegendimage{famPQ, only marks}\addlegendentry{PQ}
\addlegendimage{famRK, only marks}\addlegendentry{R-KMeans}
\addlegendimage{famPRK, mark options={solid, fill=hueKM, draw=white, line width=0.3pt}, only marks}\addlegendentry{PQ$\times$R-KMeans}
\addlegendimage{famRV, only marks}\addlegendentry{R-VQ}
\addlegendimage{famPRV, mark options={solid, fill=hueVQ, draw=white, line width=0.3pt}, only marks}\addlegendentry{PQ$\times$R-VQ}
\end{axis}
\end{tikzpicture}\\[2pt]
  \begin{subfigure}[t]{0.32\linewidth}
    \centering
    \resizebox{\linewidth}{!}{\begin{tikzpicture}
\begin{axis}[docid axis, xmode=log, x dir=reverse, xmin=0.01438, xmax=0.7605, xtick={0.6,0.2,0.1,0.02}, xticklabels={$0.4$,$0.8$,$0.9$,$0.98$}, xlabel={Uniqueness Ratio $\mathcal{U}$}, ylabel={MRR@10 (\%)}, ymin=0, ymax=50, ytick distance=10]
\draw[inkMuted, line width=0.4pt, dash pattern=on 1.5pt off 1pt] (axis cs:0.1,0) -- (axis cs:0.1,50);
\addplot+[famPQ, only marks, mark options={solid, fill=white, line width=0.7pt}] coordinates {(0.585,14.77) (0.2051,31.2) (0.1862,33.54) (0.1782,32.55)}; %
\addplot+[famRK, only marks, mark options={solid, fill=white, line width=0.7pt}] coordinates {(0.4705,26.7)}; %
\addplot+[famPRV, mark options={solid, fill=hueVQ, draw=white, line width=0.3pt}, only marks, mark options={solid, fill=white, line width=0.7pt}] coordinates {(0.3432,29.14)}; %
\addplot+[famPQ, only marks] coordinates {(0.0255,43.11) (0.0269,44.16) (0.0292,44.17) (0.0199,43.67) (0.0206,43.84) (0.021,43.82)}; %
\addplot+[famRK, only marks] coordinates {(0.024,43.72) (0.0223,43.8) (0.024,43.98) (0.0193,45.31) (0.0191,44.47) (0.0193,45.47) (0.0189,44.41) (0.0187,43.43) (0.0189,44.33)}; %
\addplot+[famPRK, mark options={solid, fill=hueKM, draw=white, line width=0.3pt}, only marks] coordinates {(0.0271,45.69) (0.0194,46.3) (0.0189,45.36) (0.0327,43.96) (0.0195,44.64) (0.0189,44.3) (0.0192,44.61) (0.0202,43.05) (0.0195,42.75)}; %
\addplot+[famRV, only marks] coordinates {(0.0506,44.02) (0.022,45.23) (0.0213,46.1) (0.0209,44.78) (0.0197,45.04) (0.0198,45.75) (0.0197,44.61) (0.0193,45.56) (0.0189,45.27)}; %
\addplot+[famPRV, mark options={solid, fill=hueVQ, draw=white, line width=0.3pt}, only marks] coordinates {(0.0262,45.54) (0.0207,47.17) (0.02,44.68) (0.0715,43.02) (0.0203,45.56) (0.0197,45.41) (0.0307,44.77) (0.0198,45.08) (0.0223,44.75) (0.0202,44.19)}; %
\end{axis}
\end{tikzpicture}}
    \caption{MS MARCO}
    \label{fig:mrr-vs-uniqueness-msmarco}
  \end{subfigure}
  \hfill
  \begin{subfigure}[t]{0.32\linewidth}
    \centering
    \resizebox{\linewidth}{!}{\begin{tikzpicture}
\begin{axis}[docid axis, xmode=log, x dir=reverse, xmin=0.00023, xmax=0.8207, xtick={0.1,0.01,0.001}, xticklabels={$0.9$,$0.99$,$0.999$}, xlabel={Uniqueness Ratio $\mathcal{U}$}, ylabel={MRR@10 (\%)}, ymin=0, ymax=60, ytick distance=10]
\draw[inkMuted, line width=0.4pt, dash pattern=on 1.5pt off 1pt] (axis cs:0.1,0) -- (axis cs:0.1,60);
\addplot+[famPQ, only marks, mark options={solid, fill=white, line width=0.7pt}] coordinates {(0.1787,45.94) (0.1511,47.23) (0.1315,48.89)}; %
\addplot+[famPRK, mark options={solid, fill=hueKM, draw=white, line width=0.3pt}, only marks, mark options={solid, fill=white, line width=0.7pt}] coordinates {(0.6313,26.73)}; %
\addplot+[famPRV, mark options={solid, fill=hueVQ, draw=white, line width=0.3pt}, only marks, mark options={solid, fill=white, line width=0.7pt}] coordinates {(0.2099,46.05)}; %
\addplot+[famPQ, only marks] coordinates {(0.0128,52.76) (0.0139,54.49) (0.0138,53.67) (0.0022,53.55) (0.0028,53.51)}; %
\addplot+[famRK, only marks] coordinates {(0.005,54.43) (0.0025,54.01) (0.0054,52.77) (0.0008,54.7) (0.0007,54.01) (0.0009,53.17) (0.0004,53.76) (0.0003,53.45) (0.0004,52.99)}; %
\addplot+[famPRK, mark options={solid, fill=hueKM, draw=white, line width=0.3pt}, only marks] coordinates {(0.0105,53.65) (0.001,54.45) (0.0005,53.99) (0.0093,53.4) (0.0015,54.66) (0.0006,54) (0.0007,54.67) (0.0015,55.04) (0.0007,53.2)}; %
\addplot+[famRV, only marks] coordinates {(0.0024,55.34) (0.002,55.14) (0.002,56.64) (0.0008,54.96) (0.0008,54.57) (0.001,55.14) (0.0003,52.68) (0.0004,52.34) (0.0005,53.23)}; %
\addplot+[famPRV, mark options={solid, fill=hueVQ, draw=white, line width=0.3pt}, only marks] coordinates {(0.0039,54.99) (0.0015,54.3) (0.0227,54.35) (0.0059,54.69)}; %
\end{axis}
\end{tikzpicture}}
    \caption{NQ}
    \label{fig:mrr-vs-uniqueness-nq}
  \end{subfigure}
  \hfill
  \begin{subfigure}[t]{0.32\linewidth}
    \centering
    \resizebox{\linewidth}{!}{\begin{tikzpicture}
\begin{axis}[docid axis, xlabel={DocID length $M$}, ylabel={Uniqueness Ratio $\mathcal{U}$}, xmode=log, log basis x=2, xmin=3.4, xmax=28, xtick={4,8,16,24}, xticklabels={4,8,16,24}, log ticks with fixed point, ymin=0.337, ymax=1.020]
\addplot+[famPQ, dsMS, error bars/.cd, y dir=both, y explicit] coordinates {(4,0.415) += (0,0) -= (0,0) (8,0.8102) += (0,0.0116) -= (0,0.0153) (16,0.9728) += (0,0.0017) -= (0,0.002) (24,0.9795) += (0,0.0006) -= (0,0.0005)}; %
\addplot+[famRK, dsMS, error bars/.cd, y dir=both, y explicit] coordinates {(4,0.5295) += (0,0) -= (0,0) (8,0.9766) += (0,0.00113) -= (0,0.000567) (16,0.9808) += (0,0.000133) -= (0,6.67e-05) (24,0.9812) += (0,0.000133) -= (0,6.67e-05)}; %
\addplot+[famPRK, mark options={solid, fill=hueKM, draw=white, line width=0.3pt}, dsMS, error bars/.cd, y dir=both, y explicit] coordinates {(8,0.9701) += (0,0.0028) -= (0,0.0028) (16,0.9803) += (0,0.0003) -= (0,0.0005) (24,0.9809) += (0,0.000225) -= (0,0.000375)}; %
\addplot+[famRV, dsMS, error bars/.cd, y dir=both, y explicit] coordinates {(4,0.9494) += (0,0) -= (0,0) (8,0.9786) += (0,0.0005) -= (0,0.0006) (16,0.9803) += (0,3.33e-05) -= (0,6.67e-05) (24,0.981) += (0,0.000133) -= (0,0.000267)}; %
\addplot+[famPRV, mark options={solid, fill=hueVQ, draw=white, line width=0.3pt}, dsMS, error bars/.cd, y dir=both, y explicit] coordinates {(4,0.6568) += (0,0) -= (0,0) (8,0.9511) += (0,0.0227) -= (0,0.0226) (12,0.9743) += (0,0.005) -= (0,0.005) (16,0.9791) += (0,0.000867) -= (0,0.00143) (24,0.9801) += (0,0.0002) -= (0,0.0003)}; %
\addplot+[famPQ, dsNQ, error bars/.cd, y dir=both, y explicit] coordinates {(8,0.8462) += (0,0.0223) -= (0,0.0249) (16,0.9865) += (0,0.0007) -= (0,0.0004) (24,0.9975) += (0,0.0003) -= (0,0.0003)}; %
\addplot+[famRK, dsNQ, error bars/.cd, y dir=both, y explicit] coordinates {(8,0.9957) += (0,0.0018) -= (0,0.0011) (16,0.9992) += (0,0.0001) -= (0,0.0001) (24,0.9996) += (0,6.67e-05) -= (0,3.33e-05)}; %
\addplot+[famPRK, mark options={solid, fill=hueKM, draw=white, line width=0.3pt}, dsNQ, error bars/.cd, y dir=both, y explicit] coordinates {(4,0.3687) += (0,0) -= (0,0) (8,0.9901) += (0,0.0006) -= (0,0.0006) (12,0.9977) += (0,0.00055) -= (0,0.00055) (16,0.9987) += (0,0.000333) -= (0,0.000167) (24,0.9994) += (0,0.000125) -= (0,7.5e-05)}; %
\addplot+[famRV, dsNQ, error bars/.cd, y dir=both, y explicit] coordinates {(8,0.9979) += (0,0.000133) -= (0,0.000267) (16,0.9991) += (0,6.67e-05) -= (0,0.000133) (24,0.9996) += (0,0.0001) -= (0,0.0001)}; %
\addplot+[famPRV, mark options={solid, fill=hueVQ, draw=white, line width=0.3pt}, dsNQ, error bars/.cd, y dir=both, y explicit] coordinates {(4,0.7901) += (0,0) -= (0,0) (8,0.9867) += (0,0.0094) -= (0,0.0094) (12,0.9963) += (0,0.0022) -= (0,0.0022) (16,0.9982) += (0,0.0008) -= (0,0.001)}; %
\end{axis}
\end{tikzpicture}}
    \caption{$\mathcal{U}$ vs. $M$}
    \label{fig:u-vs-m}
  \end{subfigure}

  \begin{subfigure}[t]{0.32\linewidth}
    \centering
    \resizebox{\linewidth}{!}{\begin{tikzpicture}
\begin{axis}[docid axis, xlabel={DocID length $M$}, ylabel={Shared Similarity $\mathcal{S}^{\mathrm{adapted}}$}, xmode=log, log basis x=2, xmin=3.4, xmax=28, xtick={4,8,16,24}, xticklabels={4,8,16,24}, log ticks with fixed point, ymin=0.613, ymax=0.982]
\addplot+[famPQ, dsMS, error bars/.cd, y dir=both, y explicit] coordinates {(4,0.6887) += (0,0) -= (0,0) (8,0.7126) += (0,0.0233) -= (0,0.0209) (16,0.843) += (0,0.00993) -= (0,0.0106) (24,0.9251) += (0,0.00678) -= (0,0.00767)}; %
\addplot+[famRK, dsMS, error bars/.cd, y dir=both, y explicit] coordinates {(4,0.663) += (0,0) -= (0,0) (8,0.7948) += (0,0.0107) -= (0,0.00603) (16,0.8965) += (0,0.00668) -= (0,0.00398) (24,0.9316) += (0,0.00414) -= (0,0.00219)}; %
\addplot+[famPRK, mark options={solid, fill=hueKM, draw=white, line width=0.3pt}, dsMS, error bars/.cd, y dir=both, y explicit] coordinates {(8,0.7651) += (0,0.0172) -= (0,0.0172) (16,0.875) += (0,0.0162) -= (0,0.0122) (24,0.9156) += (0,0.0116) -= (0,0.0193)}; %
\addplot+[famRV, dsMS, error bars/.cd, y dir=both, y explicit] coordinates {(4,0.7971) += (0,0) -= (0,0) (8,0.8741) += (0,0.0213) -= (0,0.0229) (16,0.9371) += (0,0.00948) -= (0,0.011) (24,0.9573) += (0,0.0064) -= (0,0.00749)}; %
\addplot+[famPRV, mark options={solid, fill=hueVQ, draw=white, line width=0.3pt}, dsMS, error bars/.cd, y dir=both, y explicit] coordinates {(4,0.7028) += (0,0) -= (0,0) (8,0.7976) += (0,0.0274) -= (0,0.0274) (12,0.8523) += (0,0.0276) -= (0,0.0276) (16,0.8917) += (0,0.024) -= (0,0.0213) (24,0.9215) += (0,0.0058) -= (0,0.00635)}; %
\addplot+[famPQ, dsNQ, error bars/.cd, y dir=both, y explicit] coordinates {(8,0.7388) += (0,0.0164) -= (0,0.0164) (16,0.8077) += (0,0.0191) -= (0,0.0161) (24,0.8802) += (0,0.0209) -= (0,0.0182)}; %
\addplot+[famRK, dsNQ, error bars/.cd, y dir=both, y explicit] coordinates {(8,0.791) += (0,0.0337) -= (0,0.0373) (16,0.8905) += (0,0.0186) -= (0,0.0214) (24,0.9265) += (0,0.0124) -= (0,0.0146)}; %
\addplot+[famPRK, mark options={solid, fill=hueKM, draw=white, line width=0.3pt}, dsNQ, error bars/.cd, y dir=both, y explicit] coordinates {(4,0.7006) += (0,0) -= (0,0) (8,0.755) += (0,0.00594) -= (0,0.00594) (12,0.8262) += (0,0.00095) -= (0,0.00095) (16,0.8584) += (0,0.0134) -= (0,0.00956) (24,0.9093) += (0,0.00352) -= (0,0.00896)}; %
\addplot+[famRV, dsNQ, error bars/.cd, y dir=both, y explicit] coordinates {(8,0.8786) += (0,0.0157) -= (0,0.0163) (16,0.9393) += (0,0.00796) -= (0,0.00965) (24,0.959) += (0,0.00636) -= (0,0.00732)}; %
\addplot+[famPRV, mark options={solid, fill=hueVQ, draw=white, line width=0.3pt}, dsNQ, error bars/.cd, y dir=both, y explicit] coordinates {(4,0.7347) += (0,0) -= (0,0) (8,0.8074) += (0,0.0281) -= (0,0.0281) (12,0.8607) += (0,0.0325) -= (0,0.0325) (16,0.8911) += (0,0.0249) -= (0,0.0317)}; %
\end{axis}
\end{tikzpicture}}
    \caption{$\mathcal{S}$ (Adapted) vs. $M$}
    \label{fig:f3-s-vs-m}
  \end{subfigure}
  \hfill
  \begin{subfigure}[t]{0.32\linewidth}
    \centering
    \resizebox{\linewidth}{!}{\begin{tikzpicture}
\begin{axis}[docid axis, xlabel={DocID length $M$}, ylabel={Shared Similarity $\mathcal{S}^{\mathrm{hier}}$}, xmode=log, log basis x=2, xmin=3.4, xmax=28, xtick={4,8,16,24}, xticklabels={4,8,16,24}, log ticks with fixed point, ymin=0.613, ymax=0.982]
\addplot+[famPQ, dsMS, error bars/.cd, y dir=both, y explicit] coordinates {(4,0.6892) += (0,0) -= (0,0) (8,0.7137) += (0,0.0264) -= (0,0.0242) (16,0.842) += (0,0.0137) -= (0,0.0145) (24,0.9265) += (0,0.00493) -= (0,0.00721)}; %
\addplot+[famRK, dsMS, error bars/.cd, y dir=both, y explicit] coordinates {(4,0.6593) += (0,0) -= (0,0) (8,0.7963) += (0,0.00907) -= (0,0.00665) (16,0.8972) += (0,0.00498) -= (0,0.00347) (24,0.9315) += (0,0.00333) -= (0,0.00231)}; %
\addplot+[famPRK, mark options={solid, fill=hueKM, draw=white, line width=0.3pt}, dsMS, error bars/.cd, y dir=both, y explicit] coordinates {(8,0.7666) += (0,0.00443) -= (0,0.00443) (16,0.8654) += (0,0.0192) -= (0,0.0308) (24,0.9044) += (0,0.0186) -= (0,0.0194)}; %
\addplot+[famRV, dsMS, error bars/.cd, y dir=both, y explicit] coordinates {(4,0.8) += (0,0) -= (0,0) (8,0.8718) += (0,0.0195) -= (0,0.0221) (16,0.9365) += (0,0.00961) -= (0,0.0113) (24,0.9578) += (0,0.0067) -= (0,0.00734)}; %
\addplot+[famPRV, mark options={solid, fill=hueVQ, draw=white, line width=0.3pt}, dsMS, error bars/.cd, y dir=both, y explicit] coordinates {(4,0.6938) += (0,0) -= (0,0) (8,0.7841) += (0,0.0307) -= (0,0.0307) (12,0.8378) += (0,0.0387) -= (0,0.0387) (16,0.8678) += (0,0.0398) -= (0,0.0396) (24,0.8958) += (0,0.0173) -= (0,0.0199)}; %
\addplot+[famPQ, dsNQ, error bars/.cd, y dir=both, y explicit] coordinates {(8,0.7352) += (0,0.0213) -= (0,0.0203) (16,0.8065) += (0,0.0206) -= (0,0.0177) (24,0.881) += (0,0.0195) -= (0,0.019)}; %
\addplot+[famRK, dsNQ, error bars/.cd, y dir=both, y explicit] coordinates {(8,0.7884) += (0,0.0331) -= (0,0.0386) (16,0.8895) += (0,0.018) -= (0,0.0215) (24,0.9259) += (0,0.012) -= (0,0.0144)}; %
\addplot+[famPRK, mark options={solid, fill=hueKM, draw=white, line width=0.3pt}, dsNQ, error bars/.cd, y dir=both, y explicit] coordinates {(4,0.7112) += (0,0) -= (0,0) (8,0.7614) += (0,0.00425) -= (0,0.00425) (12,0.8241) += (0,0.00587) -= (0,0.00587) (16,0.8554) += (0,0.0159) -= (0,0.0256) (24,0.9026) += (0,0.011) -= (0,0.0221)}; %
\addplot+[famRV, dsNQ, error bars/.cd, y dir=both, y explicit] coordinates {(8,0.8781) += (0,0.0182) -= (0,0.0176) (16,0.9386) += (0,0.00892) -= (0,0.0105) (24,0.9586) += (0,0.00574) -= (0,0.00672)}; %
\addplot+[famPRV, mark options={solid, fill=hueVQ, draw=white, line width=0.3pt}, dsNQ, error bars/.cd, y dir=both, y explicit] coordinates {(4,0.7259) += (0,0) -= (0,0) (8,0.8015) += (0,0.0309) -= (0,0.0309) (12,0.8447) += (0,0.0384) -= (0,0.0384) (16,0.8753) += (0,0.036) -= (0,0.0395)}; %
\end{axis}
\end{tikzpicture}}
    \caption{$\mathcal{S}$ (Hier) vs. $M$}
    \label{fig:f3-s-hier-vs-m}
  \end{subfigure}
  \hfill
  \begin{subfigure}[t]{0.32\linewidth}
    \centering
    \resizebox{\linewidth}{!}{\begin{tikzpicture}
\begin{axis}[docid axis, xlabel={DocID length $M$}, ylabel={Shared Similarity $\mathcal{S}^{\mathrm{parallel}}$}, xmode=log, log basis x=2, xmin=3.4, xmax=28, xtick={4,8,16,24}, xticklabels={4,8,16,24}, log ticks with fixed point, ymin=0.613, ymax=0.982]
\addplot+[famPQ, dsMS, error bars/.cd, y dir=both, y explicit] coordinates {(4,0.688) += (0,0) -= (0,0) (8,0.7103) += (0,0.0245) -= (0,0.023) (16,0.8437) += (0,0.0163) -= (0,0.0129) (24,0.9247) += (0,0.00655) -= (0,0.00859)}; %
\addplot+[famRK, dsMS, error bars/.cd, y dir=both, y explicit] coordinates {(4,0.6302) += (0,0) -= (0,0) (8,0.7425) += (0,0.024) -= (0,0.0325) (16,0.8576) += (0,0.0185) -= (0,0.0289) (24,0.8992) += (0,0.0177) -= (0,0.0234)}; %
\addplot+[famPRK, mark options={solid, fill=hueKM, draw=white, line width=0.3pt}, dsMS, error bars/.cd, y dir=both, y explicit] coordinates {(8,0.7508) += (0,0.00444) -= (0,0.00444) (16,0.8703) += (0,0.00725) -= (0,0.00942) (24,0.9184) += (0,0.0201) -= (0,0.0108)}; %
\addplot+[famRV, dsMS, error bars/.cd, y dir=both, y explicit] coordinates {(4,0.7719) += (0,0) -= (0,0) (8,0.8544) += (0,0.0216) -= (0,0.0253) (16,0.922) += (0,0.00857) -= (0,0.00667) (24,0.9488) += (0,0.00687) -= (0,0.0064)}; %
\addplot+[famPRV, mark options={solid, fill=hueVQ, draw=white, line width=0.3pt}, dsMS, error bars/.cd, y dir=both, y explicit] coordinates {(4,0.6957) += (0,0) -= (0,0) (8,0.8059) += (0,0.0317) -= (0,0.0317) (12,0.8666) += (0,0.0289) -= (0,0.0289) (16,0.9043) += (0,0.015) -= (0,0.0218) (24,0.9415) += (0,0.0057) -= (0,0.00726)}; %
\addplot+[famPQ, dsNQ, error bars/.cd, y dir=both, y explicit] coordinates {(8,0.7376) += (0,0.0198) -= (0,0.0179) (16,0.8095) += (0,0.0168) -= (0,0.0175) (24,0.8796) += (0,0.0203) -= (0,0.0185)}; %
\addplot+[famRK, dsNQ, error bars/.cd, y dir=both, y explicit] coordinates {(8,0.7161) += (0,0.0318) -= (0,0.0355) (16,0.7881) += (0,0.027) -= (0,0.024) (24,0.7796) += (0,0.0552) -= (0,0.0338)}; %
\addplot+[famPRK, mark options={solid, fill=hueKM, draw=white, line width=0.3pt}, dsNQ, error bars/.cd, y dir=both, y explicit] coordinates {(4,0.6825) += (0,0) -= (0,0) (8,0.7368) += (0,0.0197) -= (0,0.0197) (12,0.8226) += (0,0.0134) -= (0,0.0134) (16,0.8427) += (0,0.0133) -= (0,0.0161) (24,0.8996) += (0,0.0283) -= (0,0.0112)}; %
\addplot+[famRV, dsNQ, error bars/.cd, y dir=both, y explicit] coordinates {(8,0.8547) += (0,0.0167) -= (0,0.0178) (16,0.9047) += (0,0.00915) -= (0,0.00959) (24,0.8874) += (0,0.0352) -= (0,0.0325)}; %
\addplot+[famPRV, mark options={solid, fill=hueVQ, draw=white, line width=0.3pt}, dsNQ, error bars/.cd, y dir=both, y explicit] coordinates {(4,0.7285) += (0,0) -= (0,0) (8,0.8083) += (0,0.0204) -= (0,0.0204) (12,0.8658) += (0,0.0244) -= (0,0.0244) (16,0.9039) += (0,0.0188) -= (0,0.0254)}; %
\end{axis}
\end{tikzpicture}}
    \caption{$\mathcal{S}$ (Parallel) vs. $M$}
    \label{fig:f3-s-full-vs-m}
  \end{subfigure}

  \begin{subfigure}[t]{0.32\linewidth}
    \centering
    \resizebox{\linewidth}{!}{\begin{tikzpicture}
\begin{axis}[docid axis, xlabel={DocID length $M$}, ylabel={Alignment $\mathcal{A}^{\mathrm{adapted}}$}, xmode=log, log basis x=2, xmin=3.4, xmax=28, xtick={4,8,16,24}, xticklabels={4,8,16,24}, log ticks with fixed point, ymin=0.497, ymax=0.986]
\addplot+[famPQ, dsMS, error bars/.cd, y dir=both, y explicit] coordinates {(4,0.938) += (0,0) -= (0,0) (8,0.8677) += (0,0.0333) -= (0,0.0427) (16,0.7953) += (0,0.0407) -= (0,0.0333) (24,0.7423) += (0,0.0237) -= (0,0.0383)}; %
\addplot+[famRK, dsMS, error bars/.cd, y dir=both, y explicit] coordinates {(4,0.932) += (0,0) -= (0,0) (8,0.922) += (0,0.021) -= (0,0.026) (16,0.9233) += (0,0.0197) -= (0,0.0273) (24,0.9207) += (0,0.0213) -= (0,0.0247)}; %
\addplot+[famPRK, mark options={solid, fill=hueKM, draw=white, line width=0.3pt}, dsMS, error bars/.cd, y dir=both, y explicit] coordinates {(8,0.914) += (0,0) -= (0,0) (16,0.8993) += (0,0.0297) -= (0,0.0443) (24,0.871) += (0,0.048) -= (0,0.067)}; %
\addplot+[famRV, dsMS, error bars/.cd, y dir=both, y explicit] coordinates {(4,0.954) += (0,0) -= (0,0) (8,0.9423) += (0,0.0117) -= (0,0.0213) (16,0.9517) += (0,0.0123) -= (0,0.0167) (24,0.9417) += (0,0.0153) -= (0,0.0247)}; %
\addplot+[famPRV, mark options={solid, fill=hueVQ, draw=white, line width=0.3pt}, dsMS, error bars/.cd, y dir=both, y explicit] coordinates {(4,0.913) += (0,0) -= (0,0) (8,0.9255) += (0,0.0195) -= (0,0.0195) (12,0.9305) += (0,0.0185) -= (0,0.0185) (16,0.9057) += (0,0.0283) -= (0,0.0227) (24,0.8813) += (0,0.0287) -= (0,0.0383)}; %
\addplot+[famPQ, dsNQ, error bars/.cd, y dir=both, y explicit] coordinates {(8,0.8907) += (0,0.0353) -= (0,0.0297) (16,0.807) += (0,0.031) -= (0,0.046) (24,0.7563) += (0,0.0227) -= (0,0.0233)}; %
\addplot+[famRK, dsNQ, error bars/.cd, y dir=both, y explicit] coordinates {(8,0.93) += (0,0.01) -= (0,0.013) (16,0.934) += (0,0.017) -= (0,0.013) (24,0.9327) += (0,0.00733) -= (0,0.0107)}; %
\addplot+[famPRK, mark options={solid, fill=hueKM, draw=white, line width=0.3pt}, dsNQ, error bars/.cd, y dir=both, y explicit] coordinates {(4,0.919) += (0,0) -= (0,0) (8,0.914) += (0,0.002) -= (0,0.002) (12,0.899) += (0,0.014) -= (0,0.014) (16,0.8963) += (0,0.0237) -= (0,0.0333) (24,0.8942) += (0,0.0318) -= (0,0.0513)}; %
\addplot+[famRV, dsNQ, error bars/.cd, y dir=both, y explicit] coordinates {(8,0.9477) += (0,0.0133) -= (0,0.0187) (16,0.938) += (0,0.021) -= (0,0.023) (24,0.946) += (0,0.011) -= (0,0.011)}; %
\addplot+[famPRV, mark options={solid, fill=hueVQ, draw=white, line width=0.3pt}, dsNQ, error bars/.cd, y dir=both, y explicit] coordinates {(4,0.948) += (0,0) -= (0,0) (8,0.939) += (0,0.006) -= (0,0.006) (12,0.923) += (0,0.004) -= (0,0.004) (16,0.9117) += (0,0.0233) -= (0,0.0327)}; %
\end{axis}
\end{tikzpicture}}
    \caption{$\mathcal{A}$ (Adapted) vs. $M$}
    \label{fig:f3-a-vs-m}
  \end{subfigure}
  \hfill
  \begin{subfigure}[t]{0.32\linewidth}
    \centering
    \resizebox{\linewidth}{!}{\begin{tikzpicture}
\begin{axis}[docid axis, xlabel={DocID length $M$}, ylabel={Alignment $\mathcal{A}^{\mathrm{hier}}$}, xmode=log, log basis x=2, xmin=3.4, xmax=28, xtick={4,8,16,24}, xticklabels={4,8,16,24}, log ticks with fixed point, ymin=0.497, ymax=0.986]
\addplot+[famPQ, dsMS, error bars/.cd, y dir=both, y explicit] coordinates {(4,0.952) += (0,0) -= (0,0) (8,0.9037) += (0,0.0303) -= (0,0.0367) (16,0.8287) += (0,0.0393) -= (0,0.0437) (24,0.7413) += (0,0.0267) -= (0,0.0253)}; %
\addplot+[famRK, dsMS, error bars/.cd, y dir=both, y explicit] coordinates {(4,0.932) += (0,0) -= (0,0) (8,0.922) += (0,0.021) -= (0,0.026) (16,0.9233) += (0,0.0197) -= (0,0.0273) (24,0.9207) += (0,0.0213) -= (0,0.0247)}; %
\addplot+[famPRK, mark options={solid, fill=hueKM, draw=white, line width=0.3pt}, dsMS, error bars/.cd, y dir=both, y explicit] coordinates {(8,0.922) += (0,0.012) -= (0,0.012) (16,0.9053) += (0,0.0257) -= (0,0.0463) (24,0.8965) += (0,0.0365) -= (0,0.0365)}; %
\addplot+[famRV, dsMS, error bars/.cd, y dir=both, y explicit] coordinates {(4,0.954) += (0,0) -= (0,0) (8,0.9423) += (0,0.0117) -= (0,0.0213) (16,0.9517) += (0,0.0123) -= (0,0.0167) (24,0.9417) += (0,0.0153) -= (0,0.0247)}; %
\addplot+[famPRV, mark options={solid, fill=hueVQ, draw=white, line width=0.3pt}, dsMS, error bars/.cd, y dir=both, y explicit] coordinates {(4,0.93) += (0,0) -= (0,0) (8,0.935) += (0,0.009) -= (0,0.009) (12,0.9255) += (0,0.0205) -= (0,0.0205) (16,0.9257) += (0,0.0203) -= (0,0.0237) (24,0.9123) += (0,0.0237) -= (0,0.0363)}; %
\addplot+[famPQ, dsNQ, error bars/.cd, y dir=both, y explicit] coordinates {(8,0.912) += (0,0.024) -= (0,0.04) (16,0.838) += (0,0.008) -= (0,0.014) (24,0.785) += (0,0.023) -= (0,0.019)}; %
\addplot+[famRK, dsNQ, error bars/.cd, y dir=both, y explicit] coordinates {(8,0.93) += (0,0.01) -= (0,0.013) (16,0.934) += (0,0.017) -= (0,0.013) (24,0.9327) += (0,0.00733) -= (0,0.0107)}; %
\addplot+[famPRK, mark options={solid, fill=hueKM, draw=white, line width=0.3pt}, dsNQ, error bars/.cd, y dir=both, y explicit] coordinates {(4,0.933) += (0,0) -= (0,0) (8,0.9285) += (0,0.0025) -= (0,0.0025) (12,0.922) += (0,0.007) -= (0,0.007) (16,0.9167) += (0,0.0223) -= (0,0.0227) (24,0.9138) += (0,0.0202) -= (0,0.0328)}; %
\addplot+[famRV, dsNQ, error bars/.cd, y dir=both, y explicit] coordinates {(8,0.9477) += (0,0.0133) -= (0,0.0187) (16,0.938) += (0,0.021) -= (0,0.023) (24,0.946) += (0,0.011) -= (0,0.011)}; %
\addplot+[famPRV, mark options={solid, fill=hueVQ, draw=white, line width=0.3pt}, dsNQ, error bars/.cd, y dir=both, y explicit] coordinates {(4,0.935) += (0,0) -= (0,0) (8,0.942) += (0,0.001) -= (0,0.001) (12,0.938) += (0,0) -= (0,0) (16,0.9337) += (0,0.0143) -= (0,0.0277)}; %
\end{axis}
\end{tikzpicture}}
    \caption{$\mathcal{A}$ (Hier) vs. $M$}
    \label{fig:f3-a-hier-vs-m}
  \end{subfigure}
  \hfill
  \begin{subfigure}[t]{0.32\linewidth}
    \centering
    \resizebox{\linewidth}{!}{\begin{tikzpicture}
\begin{axis}[docid axis, xlabel={DocID length $M$}, ylabel={Alignment $\mathcal{A}^{\mathrm{parallel}}$}, xmode=log, log basis x=2, xmin=3.4, xmax=28, xtick={4,8,16,24}, xticklabels={4,8,16,24}, log ticks with fixed point, ymin=0.497, ymax=0.986]
\addplot+[famPQ, dsMS, error bars/.cd, y dir=both, y explicit] coordinates {(4,0.938) += (0,0) -= (0,0) (8,0.8677) += (0,0.0333) -= (0,0.0427) (16,0.7953) += (0,0.0407) -= (0,0.0333) (24,0.7423) += (0,0.0237) -= (0,0.0383)}; %
\addplot+[famRK, dsMS, error bars/.cd, y dir=both, y explicit] coordinates {(4,0.519) += (0,0) -= (0,0) (8,0.5787) += (0,0.0163) -= (0,0.0177) (16,0.572) += (0,0.02) -= (0,0.026) (24,0.5697) += (0,0.0223) -= (0,0.0177)}; %
\addplot+[famPRK, mark options={solid, fill=hueKM, draw=white, line width=0.3pt}, dsMS, error bars/.cd, y dir=both, y explicit] coordinates {(8,0.5685) += (0,0.0085) -= (0,0.0085) (16,0.574) += (0,0.021) -= (0,0.029) (24,0.6072) += (0,0.0708) -= (0,0.0323)}; %
\addplot+[famRV, dsMS, error bars/.cd, y dir=both, y explicit] coordinates {(4,0.808) += (0,0) -= (0,0) (8,0.7237) += (0,0.0223) -= (0,0.0317) (16,0.657) += (0,0.018) -= (0,0.011) (24,0.6207) += (0,0.0153) -= (0,0.0107)}; %
\addplot+[famPRV, mark options={solid, fill=hueVQ, draw=white, line width=0.3pt}, dsMS, error bars/.cd, y dir=both, y explicit] coordinates {(4,0.847) += (0,0) -= (0,0) (8,0.7925) += (0,0.0345) -= (0,0.0345) (12,0.778) += (0,0.026) -= (0,0.026) (16,0.7253) += (0,0.0447) -= (0,0.0643) (24,0.704) += (0,0.035) -= (0,0.031)}; %
\addplot+[famPQ, dsNQ, error bars/.cd, y dir=both, y explicit] coordinates {(8,0.8907) += (0,0.0353) -= (0,0.0297) (16,0.807) += (0,0.031) -= (0,0.046) (24,0.7563) += (0,0.0227) -= (0,0.0233)}; %
\addplot+[famRK, dsNQ, error bars/.cd, y dir=both, y explicit] coordinates {(8,0.5557) += (0,0.0123) -= (0,0.0167) (16,0.5777) += (0,0.00333) -= (0,0.00467) (24,0.5797) += (0,0.0163) -= (0,0.0127)}; %
\addplot+[famPRK, mark options={solid, fill=hueKM, draw=white, line width=0.3pt}, dsNQ, error bars/.cd, y dir=both, y explicit] coordinates {(4,0.717) += (0,0) -= (0,0) (8,0.5935) += (0,0.0165) -= (0,0.0165) (12,0.5695) += (0,0.0015) -= (0,0.0015) (16,0.5687) += (0,0.00833) -= (0,0.00767) (24,0.5863) += (0,0.0797) -= (0,0.0312)}; %
\addplot+[famRV, dsNQ, error bars/.cd, y dir=both, y explicit] coordinates {(8,0.7157) += (0,0.0243) -= (0,0.0127) (16,0.654) += (0,0.021) -= (0,0.028) (24,0.6243) += (0,0.00467) -= (0,0.00933)}; %
\addplot+[famPRV, mark options={solid, fill=hueVQ, draw=white, line width=0.3pt}, dsNQ, error bars/.cd, y dir=both, y explicit] coordinates {(4,0.837) += (0,0) -= (0,0) (8,0.7755) += (0,0.0355) -= (0,0.0355) (12,0.7475) += (0,0.0625) -= (0,0.0625) (16,0.7033) += (0,0.0587) -= (0,0.0403)}; %
\end{axis}
\end{tikzpicture}}
    \caption{$\mathcal{A}$ (Parallel) vs. $M$}
    \label{fig:f3-a-full-vs-m}
  \end{subfigure}
  \caption{Uniqueness Ratio $\mathcal{U}$, Shared Semantic Similarity $\mathcal{S}$ and Contrastive Alignment Preservation $\mathcal{A}$. (a) and (b) Uniqueness $\mathcal{U}$ vs. MRR@10 on MS MARCO and NQ, one marker per configuration (color and shape: family), hollow: $\mathcal{U} < 0.9$, measured or implied by $V^M < 0.9N$ (at most $V^M$ distinct DocIDs); $\mathcal{U}$ axis stretched toward 1 (log scale of the collision rate), dashed line: $\mathcal{U}=0.9$. $R^2$ of MRR@10 on the metric, MS MARCO / NQ: $\mathcal{U}$ $0.94$ / $0.97$. 
  (c) to (i): Impact of DocID length $M$ and evaluation formulation on Uniqueness ($\mathcal{U}$), Shared Similarity ($\mathcal{S}$) and Alignment ($\mathcal{A}$) MS MARCO: full lines, NQ: dashed lines. 
  (d) to (f): $\mathcal{S}$, (g) to (i): $\mathcal{A}$, in their Adapted (\Cref{subsec:shared-similarity}), Hier and Full variants}
  \label{fig:metrics_length_analysis}
  \label{fig:results-uniqueness}
  \label{fig:f3-results-as-vs-m}
\end{figure*}
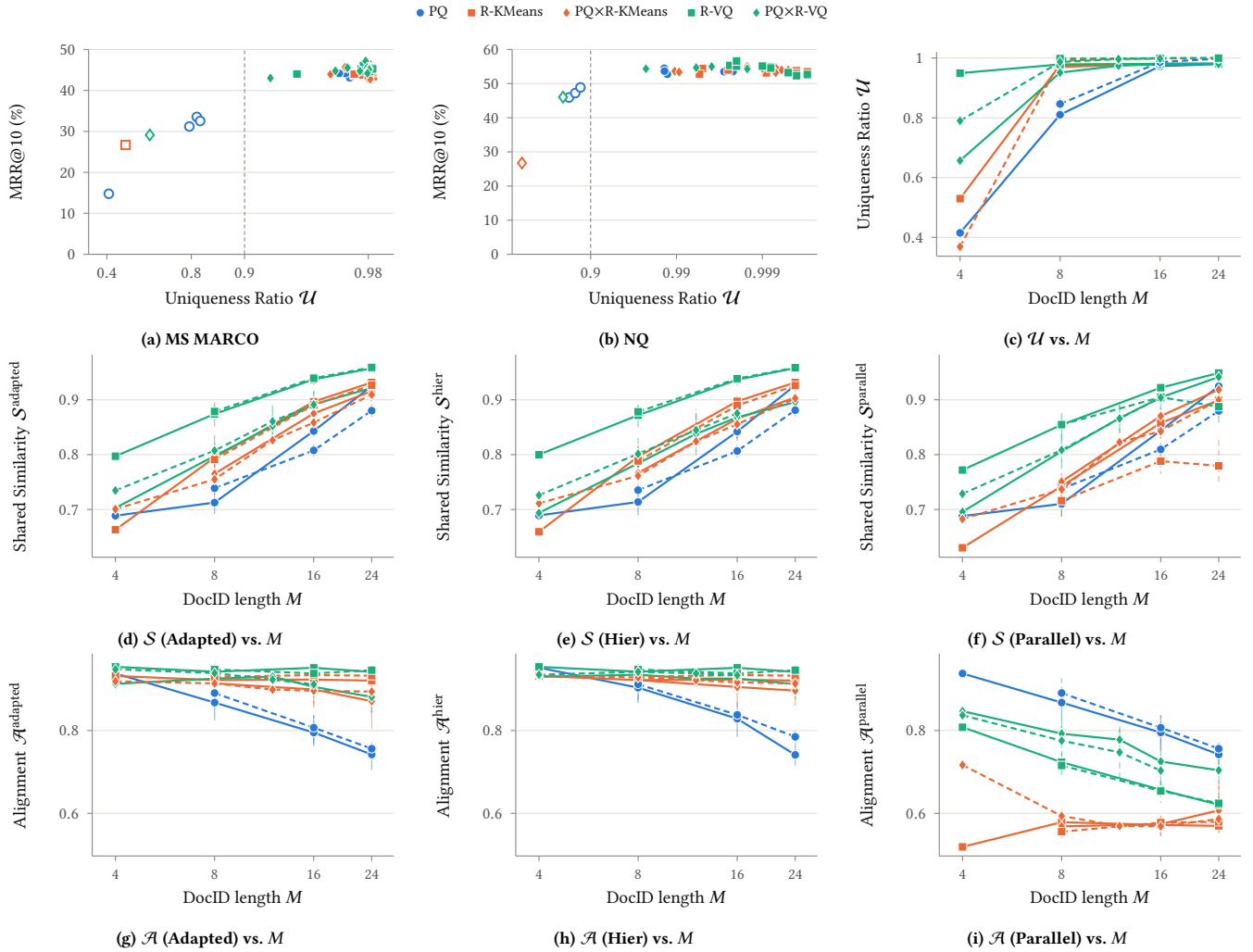

\paragraph{Uniqueness Ratio} As collisions cap recall, downstream MRR@10 collapses when $\mathcal{U}$ drops below $0.9$ (\Cref{fig:mrr-vs-uniqueness-msmarco,fig:mrr-vs-uniqueness-nq}; $R^2 = 0.94$ and $0.96$). The ceiling $\mathcal{U}_{max}$ is $0.9819$ on MS MARCO.

Evaluating uniqueness across sequence lengths $M$ reveals contrasting resilience among encoding schemes (\Cref{fig:u-vs-m}). PQ suffers from catastrophic collisions under tight budgets (from $C=8$) as parallel chunks run out of capacity. R-KMeans maintains high uniqueness down to $L=8$ before degrading at $L=4$. Conversely, R-VQ remains robust even at $L=4$, consistent with its more uniform codebook usage (see Normalized Entropy). Similar trends hold on NQ, where the smaller corpus delays collisions.

These results highlight a fundamental insight for Generative Retrieval (RQ3): \textbf{the Uniqueness Ratio $\mathcal{U}$ is the gate a DocID space must pass; beyond it, the $60$ (MS MARCO) and $48$ (NQ) spaces with $\mathcal{U} \ge 0.9$ lie within $4.0$ and $4.3$ MRR@10 points of each other.}

\begin{table}[!hp]
    \centering
    \begin{adjustbox}{max width=\columnwidth}
\begin{tabular}{l ccccc}
         \toprule
         \textbf{Dataset} & \textbf{PQ} & \textbf{R-Kmeans} & \textbf{R-VQ} & \textbf{PR-Kmeans} & \textbf{PR-VQ} \\
         \midrule
         MS MARCO & 99.51 & 75.55 & 97.95 & 74.29 & 98.40\\
         NQ & 99.48 & 69.86 & 94.21 & 70.71 & 94.88 \\
         \bottomrule
    \end{tabular}
\end{adjustbox}
    \caption{Mean entropy $\mathcal{H}$ per encoding method}
    \label{tab:results-entropy}
\end{table}

\paragraph{Normalized Entropy} Sequential residual $k$-means is the only pure scheme that uses its codes unevenly: on both MS MARCO and NQ, PQ-KMeans and RQ-VAE reach $\mathcal{H} \ge 0.92$ in every configuration, against $0.62$--$0.85$ for RQ-KMeans (\Cref{tab:results-entropy}). $\mathcal{H}$ characterizes codebook usage on its own terms: collapsed PQ-KMeans spaces keep $\mathcal{H} \ge 0.99$, and RQ-KMeans stays competitive with uneven codes. Among residual schemes, even usage goes with uniqueness under short DocIDs: at $L=4$, RQ-VAE keeps $\mathcal{U}=0.95$ where RQ-KMeans falls to $0.53$ (\Cref{fig:u-vs-m}). 

This structural disparity provides a key diagnostic explanation for the downstream performance resilience of R-VQ observed under truncated identifier budgets ($M$). Because offline R-Kmeans performs greedy, layer-by-layer clustering on remaining residual error vectors, token assignments heavily concentrate on a few dominant centroids in early layers, leading to cluster collapse and severe codebook under-utilization. In contrast, joint end-to-end gradient optimization in R-VQ evenly distributes document representations across the discrete codebook space. This high information density enables R-VQ to preserve discriminative capacity and maintain robust retrieval performance even when the identifier sequence length $M$ is heavily compressed.

\paragraph{Shared Similarity ($\mathcal{S}$)} measures the average continuous embedding similarity between documents grouped under common prefix constraints. As expected, \Cref{fig:f3-s-vs-m} shows that $\mathcal{S}$ scales monotonically with sequence depth $M$ across all schemes, as deeper prefixes restrict document clusters to increasingly compact semantic neighborhoods. End-to-end trained Residual Vector Quantization (R-VQ) consistently maintains the highest local coherence, outperforming flat Product Quantization (PQ) across all sequence budgets, regardless of whether similarity is evaluated via unstructured token overlap (\textit{Parallel}), common prefix length (\textit{Hier}), or subspace-wise prefix matching (\textit{Adapted})

\paragraph{Contrastive Alignment Preservation ($\mathcal{A}$)} diagnoses whether discrete DocID distances reflect the continuous relative proximity of document triplets. 

Figures~\ref{fig:metrics_length_analysis}(g--i) highlight a strong interaction between the structural design of the DocIDs and the evaluation metric:
\begin{itemize}
    \item \textbf{Hierarchical and Hybrid metrics (\textit{Hier}, \textit{Adapted}):} Under hierarchical evaluation (Figs.~\ref{fig:metrics_length_analysis}g and \ref{fig:metrics_length_analysis}h), alignment scores remain consistently high ($\sim 0.85\text{--}0.95$) for most configurations, showing only a slight degradation as $M$ increases for PQ DocIDs. Hierarchical RQ DocIDs perform particularly well under these metrics because early code positions capture broad cluster semantics, preserving rank alignment.
    \item \textbf{Parallel metric (\textit{Parallel}):} Conversely, when evaluated with the \textit{Parallel} metric (Fig.~\ref{fig:metrics_length_analysis}i), which measures unstructured token overlap, hierarchical RQ structures experience a noticeable decrease in alignment ($\mathcal{A}^{\text{parallel}}$ dropping below $0.6$ for certain configurations). In contrast, pure Product Quantization (PQ) DocIDs maintain higher relative stability under this metric.
\end{itemize}

Importantly, this distinct divergence in behavior between parallel and hierarchical DocIDs across the variants of $\mathcal{A}$ serves as an effective diagnostic tool. It allows us to explicitly observe and quantify the intrinsic \textit{hierarchy property} of a given DocID scheme: strong performance under $\mathcal{A}^{\text{hier}}$ coupled with degradation under $\mathcal{A}^{\text{parallel}}$ confirms that semantic information is strictly concentrated in sequential prefix order rather than distributed across independent tokens.

\section{Conclusion}
We presented a systematic study of numerical DocIDs for Generative Information Retrieval. The Uniqueness Ratio, measured before training, decides whether a DocID space is usable. The $60$ (MS MARCO) and $48$ (NQ) spaces with $\mathcal{U} \ge 0.9$ lie within $4.0$ and $4.3$ MRR@10 points of each other.

These findings come from a single parameterized family that unifies PQ, R-KMeans and R-VQ and exposes the hybrid Product--Residual Quantization (PQ$\times$RQ) space. Within it, hybrids reach the best MS MARCO result ($46.3\%$ MRR@10), while R-VQ leads on Natural Questions ($56.6\%$) and is the most robust under tight sequence budgets ($M \le 8$). In practice, $\mathcal{U}$ should be checked before training, and R-VQ preferred when DocIDs must be short.

\paragraph{Limitations and future work.}\label{sec:limitations} To isolate the DocID space, we fix the rest of the pipeline: a \texttt{t5-small} model trained without DDRO's pairwise relevance optimization~\cite{Mekonnen_2025} and decoded by prefix-constrained beam search~\cite{decao2021autoregressiveentityretrieval,wu2025constrained} rather than look-ahead~\cite{zeng2024planning}; DocIDs quantized post hoc from a frozen retrieval encoder, not learned jointly with the decoder~\cite{sun2023learningtokenizegenerativeretrieval}; collisions left unresolved, without suffix tokens or reassignment~\cite{zhang2025purelysemanticindexingllmbased}; static corpora of at most 300K documents, while generative retrieval degrades at millions of passages~\cite{pradeep2023doesgenerativeretrievalscale} and new documents require re-quantization~\cite{zhang2025replication}. $\mathcal{U}$ flags collapse ($\mathcal{U} < 0.9$), but no intrinsic metric explains MRR@10 among usable spaces ($R^2 \le 0.30$). Future work follows from these results.
\bibliographystyle{ACM-Reference-Format}
\bibliography{sample-base}

\end{document}